\documentclass[epjST,final]{svjour}[]
\usepackage{amsmath,amsfonts,amssymb,lmodern}

\usepackage[usenames,dvipsnames]{color}

\usepackage{here,subfigure}

\usepackage{cite}

\usepackage[rightcaption]{sidecap}


\usepackage{undertilde}
\usepackage{dsfont}

\newcommand{\rmd}{\mbox{d}}

\newcommand{\mean}[1]{\left < #1 \right>}
\newcommand{\abs}[1]{\left | #1 \right |}

\newcommand{\fourier}[2]{{\hat{#1}}_{#2}}
\newcommand{\nabcp}{\utilde{\nabla}}

\newcommand{\bi}[1]{\mathbf{ #1 }}

\usepackage{graphicx}

\usepackage{pdfcomment,ulem}

\makeatletter
\newcommand{\annotate}[2][]{%
\pdfstringdef\x@title{#1}%
\edef\r{\string\r}%
\pdfstringdef\x@contents{#2}%
\pdfannot
width 2\baselineskip
height 2\baselineskip
depth 0pt
{
/Subtype /Text
/T (\x@title)
/Contents (\x@contents)
}%
}
\makeatother

\begin{document}
\title{Pattern formation in active particle systems due to competing alignment interactions}
\author{Robert Gro\ss{}mann\inst{1}\fnmsep\thanks{\email{grossmann@physik.hu-berlin.de}} \and Pawel
Romanczuk\inst{2,3}\fnmsep\thanks{\email{pawelr@princeton.edu}}
\and Markus B\"ar\inst{1} \and Lutz Schimansky-Geier\inst{4} }
\institute{Physikalisch-Technische Bundesanstalt, Abbestr. 2-12, 10587 Berlin, Germany \and Department of Ecology and Evolutionary Biology, Princeton University, Princeton, NJ 80544, USA \and Thaer Institute, Humboldt-Universit{\"at} zu Berlin, 10099 Berlin, Germany
\and Department
of Physics, Humboldt-Universit{\"at} zu Berlin, Newtonstr. 15, 12489 Berlin, Germany} 
\abstract{Recently, we proposed a self-propelled particle model with competing alignment interactions: nearby
particles tend to align their velocities whereas they anti-align their direction of motion with particles which are
further away [R. Gro\ss{}mann et al., Phys. Rev. Lett. \textbf{113}, 258104 (2014)]. 
%
%
Here, we extend our previous numerical analysis of the high density regime considering low particle densities too. 
We report on the emergence of various macroscopic patterns such as vortex arrays, mesoscale turbulence as well as the 
formation of polar clusters, polar bands and nematically ordered states.
Furthermore, we study analytically the instabilities leading to pattern formation in mean-field approximation. We argue
that these instabilities are well described by a reduced set of hydrodynamic 
equations in the limit of high density. } 
\maketitle
\section{Introduction}
\label{sec:intro}

The study of active matter is one of the main topics in modern classical statistical
mechanics \cite{toner_hydrodynamics_2005,romanczuk_active_2012,
vicsek_collective_2012,marchetti_hydrodynamics_2013,
menzel_tuned_2015}. Active matter systems are composed of self-driven entities, so called active particles, which take
up energy from their environment in order to move actively in a dissipative medium \cite{romanczuk_active_2012}. Due to
the complex interplay of self-propulsion and dissipation, the dynamics of active matter systems is inherently out
of thermodynamic equilibrium. Thus, complex spatio-temporal patterns may emerge in interacting active particle
systems. There are numerous experimental realizations of active matter, for example driven filaments on the
molecular scale \cite{schaller_polar_2010}, colloidal systems \cite{theurkauff_dynamic_2012,pohl_dynamic_2014},
bacterial colonies \cite{czirok_formation_1996,sokolov_concentration_2007,peruani_collective_2012,
dombrowski_self-concentration_2004,
sokolov_enhanced_2009,cisneros_fluid_2010,
wensink_meso-scale_2012,zhang_swarming_2009,liu_multifractal_2012} as well as macroscopic systems like flocking birds
\cite{ballerini_interaction_2008} and schooling fish \cite{katz_inferring_2011}.

The experimental study of bacterial colonies has become increasingly important for the understanding of active
matter. On the one hand, the quantitative analysis of fascinating phenomena like clustering
\cite{peruani_collective_2012} and vortex formation \cite{sumino_large-scale_2012} provide an excellent opportunity to
test theoretical predictions, and, on the other hand, experiments often reveal unforeseen novel phases of active
matter. 
Recently, a number of experiments reported the emergence of dynamic vortex patterns in dense
suspensions of \textit{Bacillus subtilis}
\cite{dombrowski_self-concentration_2004,sokolov_enhanced_2009,cisneros_fluid_2010,
wensink_meso-scale_2012}, \textit{Paenibacillus dendritiformis} \cite{zhang_swarming_2009}
as well as \textit{Escherichia coli} \cite{liu_multifractal_2012}. In 
particular, it was observed that bacteria locally align their body axis. 
However, dynamic vortices (defects) in the velocity field appeared at larger 
scales. Interestingly, vortices are separated by a characteristic length scale 
which is larger than the size of an individual bacterium but smaller
than the system size. This novel phenomenon was termed \textit{mesoscale turbulence}. 

Mesoscale turbulence was suggested to be described by a phenomenological hydrodynamic theory
\cite{wensink_meso-scale_2012,dunkel_fluid_2013} which is a combination of both, the incompressible Toner-Tu theory for
flocks \cite{toner_reanalysis_2012,chen_critical_2014} and the Swift-Hohenberg model \cite{Swift_hydrodynamic_1977}. The
Swift-Hohenberg equation describes instabilities at finite
wavelengths whereas the Toner-Tu theory is the basis for the description of flocking phenomena. A combination of both --
a Turing instability in the velocity field together with convective particle 
transport -- is sufficient to explain the pattern formation 
observed in bacterial suspensions \cite{dunkel_fluid_2013}. In
particular, the experimentally observed velocity correlation function was reproduced by this hydrodynamic theory.
However, the transport coefficients in the phenomenological model remain unknown constants. 

In order to gain a deeper understanding of this pattern formation process, a particle based model was suggested by us \cite{grossmann_vortex_2014}. In this model, particles interact via competing
velocity-alignment mechanisms: nearby particles tend to align their velocities whereas they anti-align their direction of
motion with particles which are further away. The idea of competing interactions goes back to Turing
\cite{turing_chemical_1952} who showed that
instabilities on finite length scales arise in chemical systems due to the interplay of two competing species: a
slowly diffusing activator and an
inhibitor which diffuses fast compared to the activator. Within our
active particle
system, \textit{activation} refers to the creation of local polar order by 
velocity-alignment which is \textit{inhibited} by an opposing
effect acting on larger length scales (anti-alignment). Similar types of 
competing interactions were studied in the context of spin
models (ferromagnetic coupling to nearest neighbors vs. anti-ferromagnetic interaction with next-nearest
neighbors \cite{krinsky_ising_1977,barber_non_1979,oitmaa_square_1981}) and the Kuramoto model for synchronization
(conformists vs. contrarians \cite{hong_kuramoto_2011}). However, note that -- in contrast to these examples --
the competing alignment interaction in our model leads to instabilities of the velocity field which is, in turn, related
to convective mass transport resulting in numerous novel pattern formation phenomena
\cite{grossmann_vortex_2014}. 

Generally, velocity-alignment interactions are understood as simple models for the interplay of various physical forces such as 
steric repulsion or hydrodynamics. The simplest active particle model with velocity-alignment 
interaction is the seminal Vicsek model \cite{vicsek_novel_1995}: particles move 
in continuous space at a constant speed and align their direction of motion with 
the local mean velocity plus some deviations (noise) in discrete time steps. 
Over the years, numerous modifications of the original Vicsek model  were 
discussed (see e.g. \cite{chate_modeling_2008}). In some studies, anti-alignment was included in different ways: for example, 
Menzel considered 
anti-alignment between two different particle populations 
\cite{menzel_collective_2012}, whereas Lobaskin \textit{et al.} studied a model 
where particles do only align if their velocity are already aligned to a 
certain degree \cite{lobaskin_collective_2013}. For an overview 
on pattern formation in self-propelled particle systems, we refer the interested 
reader to a number of comprehensive review articles 
\cite{vicsek_collective_2012,romanczuk_active_2012,marchetti_hydrodynamics_2013, 
menzel_tuned_2015}.


In this work, we address the following questions. In section \ref{sec:model}, we 
introduce our active particle model with competing alignment interaction in 
detail. In the subsequent section, we report the
results of numerical simulations and discuss the collective dynamics exhibited by the system. Whereas we focused on
the high density limit in \cite{grossmann_vortex_2014}, we study the phase space in detail including low
and high densities here. 
In section \ref{sec:theory}, we derive hydrodynamic equations in the mean-field 
limit allowing us to understand the observed pattern formation 
qualitatively and to link our model to previously discussed phenomenological 
theories. Moreover, we estimate the quality of the approximations involved 
in the coarse-graining process and discuss their range of
applicability. 
%
%

\section{Self-propelled particle model}
\label{sec:model}

We study an ensemble of $N$ interacting self-propelled particles (SPPs) in two spatial dimensions. We assume that
particles move at constant speed which is justified if all processes related to the self-propulsion
mechanism occur on negligibly small timescales. The generic dynamics of interacting SPPs is determined by the following
set of stochastic differential equations\footnote{We work in natural units such that the speed, the
mass of particles and the anti-alignment range are equal one. }
 \begin{align}
 \label{eqn:microscopic:model}
   \frac{\rmd \bi{r}_i(t)}{\rmd t} \, = \begin{pmatrix}
                                          \cos \varphi_i \\ \sin \varphi_i
                                        \end{pmatrix} \!,  \quad 
   \frac{\rmd \varphi_i(t)}{\rmd t} = \sqrt{2 D_{\varphi}} \, \xi_i(t) + \sum_{j\neq i}^N
\mathfrak{f}_\varphi(\bi{r}_{ji},\varphi_i,\varphi_j), 
  \end{align}
where we denote the spatial position of the $i$ths particle at time $t$ by $\vec{r}_i(t)$ and its direction of motion
by $\varphi_i(t)$. The first term on the right-hand side of the angular dynamics represents noise indicating the random
reorientation of particles due to spatial inhomogeneities or fluctuations of the
driving force, whereas the second term accounts for binary interactions among the particles. 

In the absence of inter-particle coupling, a SPP performs a persistent random 
walk: a particle moves ballistically at short timescales, whereas the motion 
is diffusive on large timescales due to
angular fluctuations. The details of this diffusion process depend on the properties of the
fluctuations \cite{mikhailov_self_1997,weber_active_2011,weber_active_2012}. Here, we consider the simplest
case of Gaussian white noise with zero mean, $\mean{ \xi_i(t)} = 0$,
and temporal $\delta$-correlations:~$\mean{ \xi_i(t) \xi_j(t')} = \delta_{ij} \, \delta (t - t')$.  
The parameter $D_\varphi$ parametrizes the intensity of the angular fluctuations. 

The binary interaction $\mathfrak{f}_\varphi(\bi{r}_{ji},\varphi_i,\varphi_j)$ generally depends on the relative
coordinate
$\vec{r}_{ji} = \vec{r}_j - \vec{r}_i$ of the particles $i$ and $j$ as well as on their individual orientations. Here,
we consider the interaction scheme proposed recently by the authors in \cite{grossmann_vortex_2014} including the
following physical effects: 
\begin{enumerate}
 \item[(i)$\,\,$] Due to the finite size of particles, these objects repel each other if they come closely
together. \\
 $\Rightarrow$ short-range repulsion
 \item[(ii)$\,$] In general, short-ranged anisotropic steric effects in combination with self-propulsion will
lead to local orientational ordering. Here, we consider polar alignment: \\
 $\Rightarrow$ velocity-alignment
 \item[(iii)] An active particle in a fluid or a polarly ordered cluster 
of swimming particles will induce hydrodynamic flows that in turn influence 
other particles. It has been shown for model swimmers that hydrodynamic 
interactions are generally anisotropic and may lead to velocity-alignment, 
anti-alignment or nematic alignment depending on the relative 
orientation and position of two swimmers as well as the swimming type (see 
e.g. \cite{baskaran_statistical_2009}). Here, we replace the very complex 
hydrodynamic interaction and take one additional effective interaction mechanism 
into account: 
%
\\
 $\Rightarrow$ anti-velocity-alignment. 
\end{enumerate}
\begin{figure}[tb]
 \begin{center}
    \includegraphics[width=0.98\textwidth]{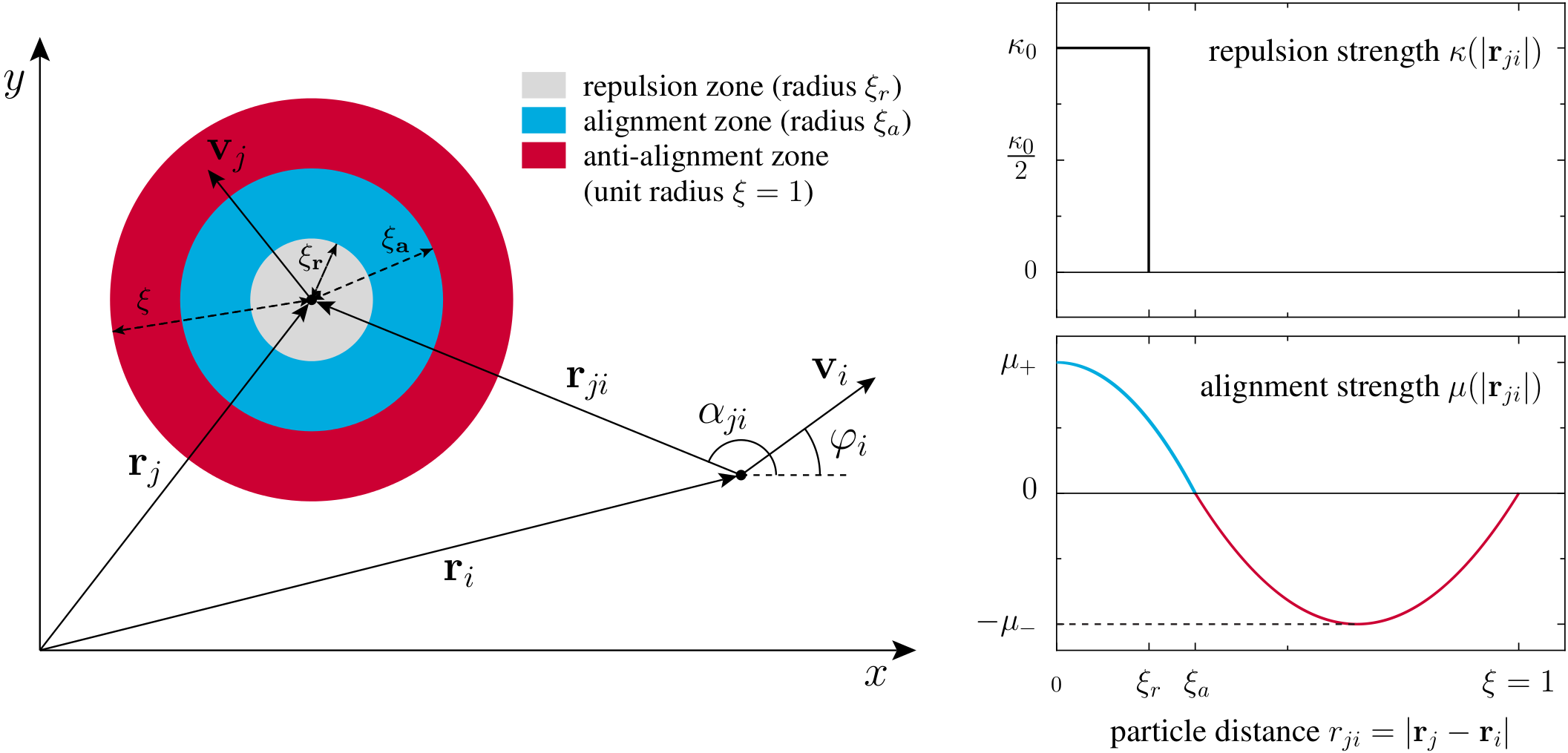}
   \caption{Visualization of geometric quantities and scheme of the pair-wise interaction. }
   \label{pic:scheme_mu}
 \end{center}
\end{figure}
We parametrize the binary interaction in \eqref{eqn:microscopic:model} by 
 \begin{align}
  \mathfrak{f}_\varphi \! \left( \bi{r}_{ji},\varphi_i,\varphi_j \right ) =  \kappa \! \left ( \abs{\bi{r}_{ji}}
\right ) \sin \left ( \varphi_i - \alpha_{ji} \right ) + \mu \! \left(\abs{\bi{r}_{ji}} \right ) \sin \left ( \varphi_j
- \varphi_i \right) \! .
\label{eq:torque}
 \end{align}
The first term represents soft-core repulsion. The angle $\alpha_{ji}=\rm{arg}(\bi{r}_{ji})$ denotes the polar
positional angle of particle $j$ in the frame of reference of particle $i$ (see also Fig. \ref{pic:scheme_mu} for
a graphical illustration). We assume a constant repulsion strength $\kappa_0$ for $\abs{\bi{r}_{ji}} < \xi_r$ and
vanishing repulsion otherwise: $\kappa \! \left ( \abs{\bi{r}_{ji}} \right ) = \kappa_0 \Theta (\xi_r -
\abs{\bi{r}_{ji}})$. 

The second term describes the competition of alignment and anti-alignment: For $\mu \!
\left(\abs{\bi{r}_{ji}} \right ) > 0$,
particles tend to align their direction of motion whereas they anti-align in the opposite case. Throughout the paper, we
will assume the following functional dependence of the alignment strength on distance
 \begin{align}
  \label{eqn:SRA:LRAA}
  \mu\! \left(\abs{\bi{r}_{ji}} \right ) = 
    \begin{cases} 
      + \mu_{+} \, A_{+}\! \left(\abs{\bi{r}_{ji}} \right )  & 0 \le \abs{\bi{r}_{ji}} \le \xi_a, \\ 
      - \mu_{-} \, A_{-}\! \left(\abs{\bi{r}_{ji}} \right )  & \xi_a \le \abs{\bi{r}_{ji}} \le 1, 
    \end{cases}
 \end{align}
%
%
where $\mu_+$ and $\mu_{-}$ are positive constants. The distance dependence of the competing interactions is determined
by the functions $A_{\pm}\! \left(\abs{\bi{r}_{ji}} \right )$. We normalize the distance dependence such that $A_{\pm}\!
\left(\abs{\bi{r}_{ji}} \right ) \in [0,1]$. Accordingly, $\mu_+$ and $\mu_-$ denote the maximal alignment and
anti-alignment strength, respectively (cf. Fig. \ref{pic:scheme_mu}). Following \cite{grossmann_vortex_2014}, we use a
piecewise quadratic distance dependence in this study: 
 \begin{align}
 \label{eqn:dist:dep:mu}
    A_{+}(r_{ji}) =  1 - \left (\frac{r_{ji}}{\xi_a}    \right)^2,  \quad
    A_{-}(r_{ji}) =  \frac{4 (r_{ji} - \xi_a)(1-r_{ji})}{(1-\xi_a)^2}.  
 \end{align}
Note that the interaction described above is strictly \textit{short-ranged} but \textit{nonlocal}. 


\section{Simulation results - pattern formation and orientational order}
\label{sec:sim}

\begin{figure}[tb]
 \begin{center}
    \includegraphics[width=0.75\textwidth]{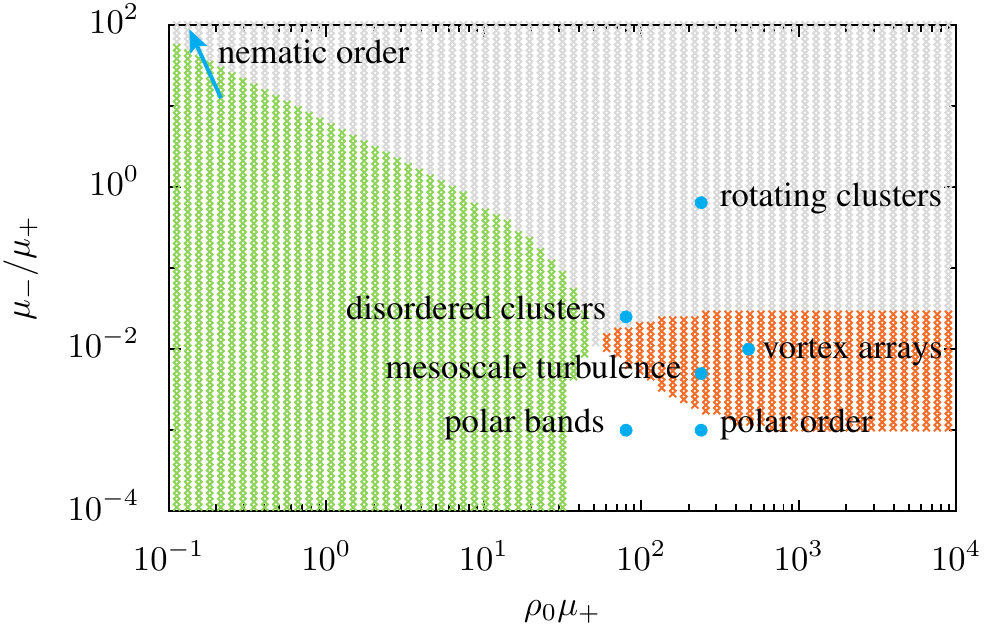}
    \caption{The figure shows the part of the 
parameter space which we study in this work. Points indicate the parameter
values for which
simulation results are shown (cf. Fig. \ref{pic_lowrho_lowmum} - Fig. \ref{pic_highrho_highmum}). The background
coloring
(green, red, gray, white) was done in accordance with the predictions of the kinetic theory: in the green region, we
expect the disordered (spatially homogeneous, isotropic) state to be stable; in the white area, we observe the emergence
of polar order (local and global);
in the gray and red colored part of the parameter space, we observe the emergence of patterns with a characteristic length scale
(see section \ref{sec:theory} for details). }
   \label{fig:PD_Sim}
 \end{center}
\end{figure}

%
%
In this section, we report on simulations of the microscopic dynamics for various parameters which were performed to
obtain a general understanding of the collective dynamics and pattern formation 
exhibited by the system. Our simulations
revealed that the system exhibits various types of spatio-temporal dynamics 
which we will discuss in the following.

We used the following fixed parameters in all simulations: repulsion range
$\xi_r=0.1$, repulsion strength $\kappa_0=10$ and alignment range $\xi_a=0.2$. 
We focus our discussion on the variation
of the alignment and anti-alignment strengths. The parameter values which we 
are going to discuss are represented in Fig. \ref{fig:PD_Sim}. 

We go beyond our previous work \cite{grossmann_vortex_2014}
and explore not only the high density regime ($\rho_0=150$) but also pattern formation at low densities ($\rho_0=20$).
The studied density values may misleadingly appear as extremely large in comparison to the values typically studied in
the Vicsek model \cite{vicsek_novel_1995,chate_collective_2008}. However, the alignment range is less than one ($\xi_a=0.2$) in our case. Hence, the mean number of
particles within a box of edge length $\xi_a$ is $\mean{n} = 0.8$ for $\rho_0 = 20$ and
$\mean{n} = 6$ for $\rho_0 = 150$. This is comparable to the regimes typically studied in the
context of the
Vicsek-model. 

All simulations were performed using a rectangular domain with
periodic boundary conditions with a linear size of $L=20=100\times\xi_a=200\times\xi_r$.  This corresponds to a rather
moderate system size with respect to the largest scale of interaction. We emphasize that the terms \textit{phase}
or \textit{state} are therefore used here to describe stereotypical patters exhibited by the system at the corresponding mesoscale
and shall not be confused with phases in the thermodynamic limit. Accordingly, \textit{global order} refers here to the
ordering of the entire simulated system and does not imply \textit{long-range 
order}. 

The initial conditions in all simulations were random positions and orientations of individual 
particles (disordered, homogeneous state). The simulation results were analyzed 
after the system had reached a stationary (dynamical) state, where no
drift in order parameters and changes in patterns were observed anymore. The 
numerical time step was chosen sufficiently small to ensure numerical 
stability: $\Delta t\leq 5\cdot10^{-3}$.

We measured polar and nematic orientational order using the corresponding order parameters
\begin{align}
	\Phi_p=\frac{1}{N} \abs{\sum_{j=1}^N \, e^{i \varphi_j}}\! , \qquad \Phi_n=\frac{1}{N} \abs{\sum_{j=1}^N \, e^{2
i \varphi_j}} \!. 
\end{align}
Furthermore, we calculated coarse-grained density $\rho_c(\vec{r},t)$ and velocity $\vec{v}_c(\vec{r},t)$ by binning of
the particles positions and velocities on a rectangular grid with the coarse-graining scale set by the linear dimension
of a grid cell: $dl=0.2$. Moreover, we calculated the momentum field via $\vec{w}_c(\vec{r},t) = \rho_c(\vec{r},t)
\vec{v}_c(\vec{r},t)$. 
%
From the coarse grained fields, we deduced the two-point autocorrelation functions
\begin{align}
 \label{eq:corrfun}
C_{\rho\rho}(\vec{r})=\overline{\rho_c(\vec{r}',t)\rho_c(\vec{r}'+\vec{r},t)}, \qquad
{C_{ww}(\vec{r})=\overline{\vec{w}_c(\vec{r}',t) \cdot \vec{w}_c(\vec{r}'+\vec{r},t)}} 
\end{align}
as well as their corresponding two-dimensional Fourier transforms $S_\rho(\vec{k})=\mathcal{F}[C_{\rho\rho}]$ and
$S_w(\vec{k})=\mathcal{F}[C_{ww}]$. In Eq. \eqref{eq:corrfun}, the overline indicates temporal averaging as well as
averaging over the reference points $\vec{r}'$. 

%
%

\subsection{Low density case ($\rho_0=20$)}

In general, no orientational or spatial order can emerge if noise is
dominant in comparison to the interaction strengths and, consequently, the system remains disordered, spatially
homogeneous and isotropic. 

\begin{figure}[p]
  \begin{center}
\includegraphics[width=0.98\textwidth]{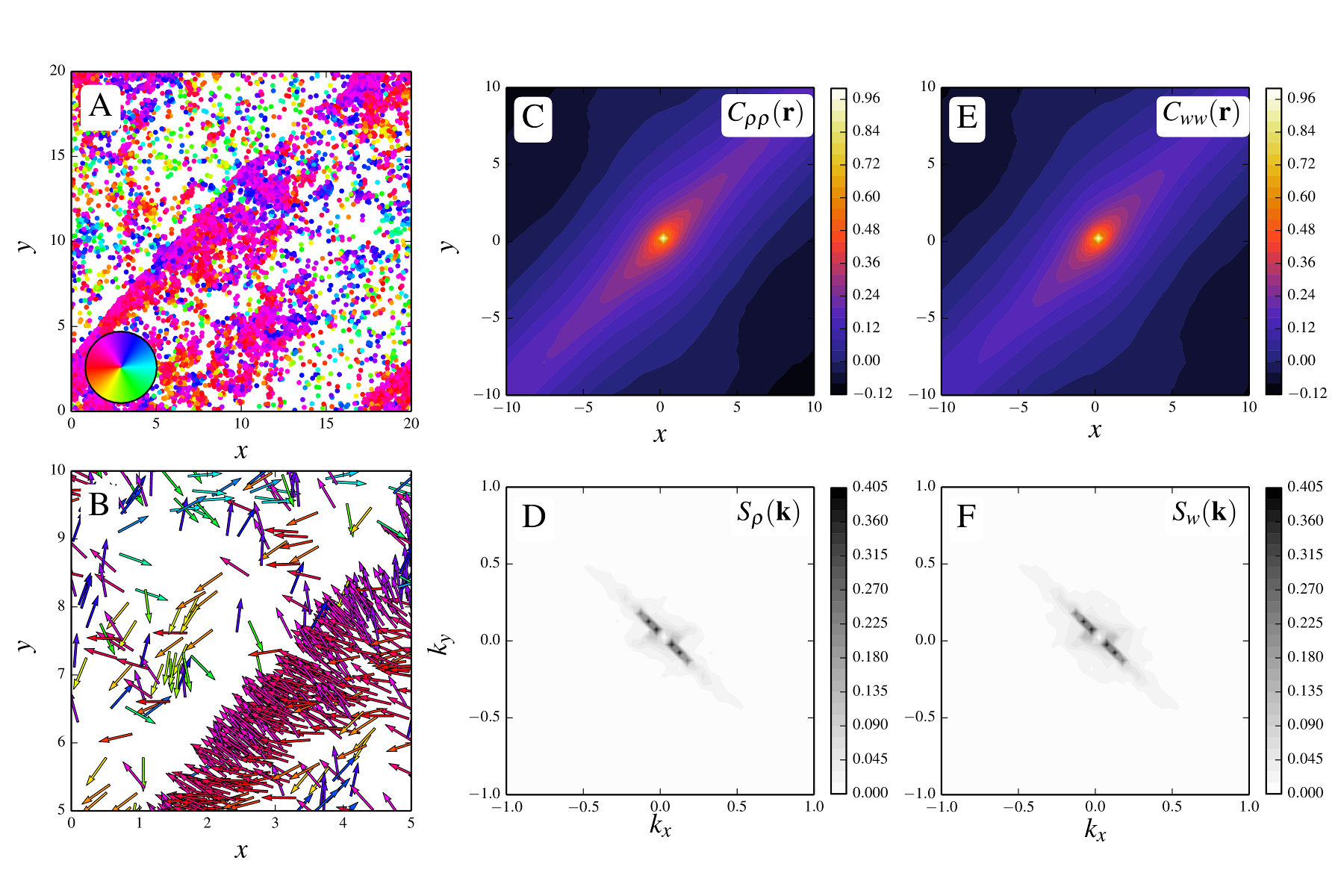}
    \caption{
\textit{Polar bands} {($\Phi_p = 0.65$ and $\Phi_n = 0.33$)} at strong alignment 
$\mu_+=4$, weak anti-alignment
$\mu_{-}=0.004$, low density $\rho_0=20$ and noise $D_\varphi=1.0$. 
({\bf A}) Global snapshot with particle positions shown by symbols and 
orientation indicated by the color according to
the inset. 
({\bf B}) Zoom into a subregion of size $5\times5$ with particle positions and 
orientations shown as vectors. 
({\bf C}) Time averaged density autocorrelation function $C_{\rho\rho}({\bf r})$ 
and ({\bf D}) its Fourier transform
$S_\rho({\bf k})$. 
({\bf E}) Time averaged momentum field autocorrelation function $C_{ww}({\bf 
r})$ and ({\bf F}) its Fourier transform
$S_w({\bf k})$. 
   \label{pic_lowrho_lowmum}
}
  \end{center}
%
  \begin{center}    
\includegraphics[width=0.98\textwidth]{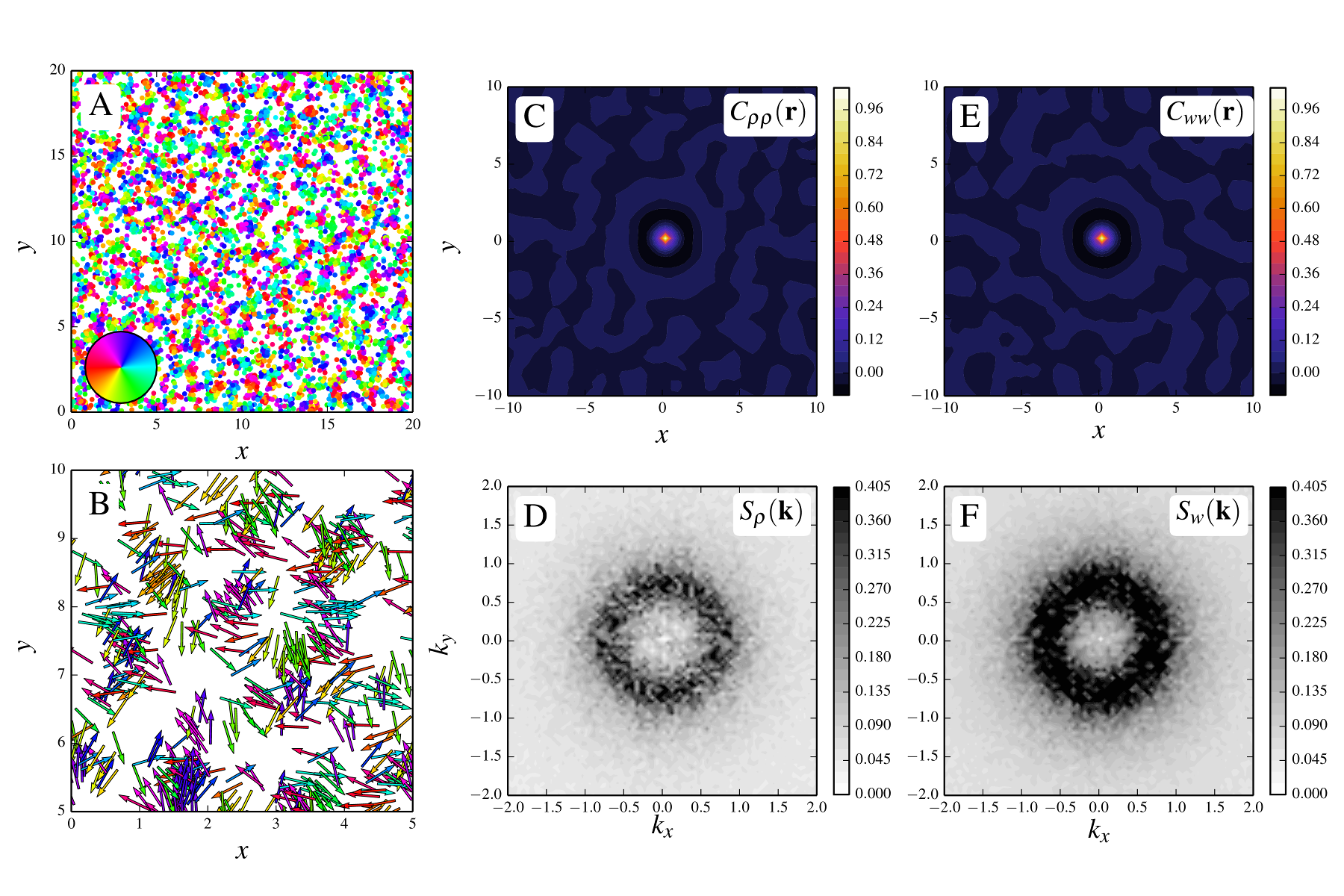}
    \caption{
\textit{Disordered clusters} {($\Phi_p = 0.021$ and $\Phi_n = 0.020$)} at strong 
alignment $\mu_+=4$, moderate
anti-alignment $\mu_-=0.1$, low density $\rho_0=20$ and noise $D_\varphi=1.0$. 
({\bf A}) Global snapshot with particle positions shown by symbols and 
orientation indicated by the color according to
the inset. 
({\bf B}) Zoom into a subregion of size $5\times5$ with particle positions and 
orientations shown as vectors. 
({\bf C}) Time averaged density autocorrelation function $C_{\rho\rho}({\bf r})$ 
and ({\bf D}) its Fourier transform
$S_\rho({\bf k})$. 
({\bf E}) Time averaged momentum field autocorrelation function $C_{ww}({\bf 
r})$ and ({\bf F}) its Fourier transform
$S_v({\bf k})$. 
   \label{pic_lowrho_modmum}
}
  \end{center}
\end{figure}

In the limit of vanishing anti-alignment ($\mu_- = 0$), the system reduces to a pure alignment model 
and the large-scale dynamics corresponds to the Vicsek model with the usual order-disorder transition 
from a disordered to a polarly ordered state at high alignment and small noise. 
In particular, we observe the emergence of large-scale travelling polar bands close to the order-disorder transition
typical for Vicsek-type models \cite{chate_collective_2008,bertin_hydrodynamic_2009}. These structures can be
clearly identified by the asymmetry of $C_{\rho\rho}$ and $C_{ww}$ reflecting an increased density and velocity
autocorrelation perpendicular to the polar orientation and low-wavenumber peaks along the polar direction in both
spectra $S_{\rho,w}$ (Fig. \ref{pic_lowrho_lowmum}). Far away from the order-disorder transition (large $\mu_+$ and low $\mu_-$), the model exhibits 
the spatially homogeneous phase with polar order described and analyzed by Toner and Tu
\cite{toner_reanalysis_2012}. 

Small levels of anti-alignment do not destroy polar order at the system sizes studied. Only if $\mu_-$ is increased
beyond a critical value, we observe the breakdown of global polar order. At low density, this breakdown is accompanied
by
the formation of small polar clusters, which are unstable and
frequently break up. The corresponding density autocorrelation and its Fourier transform are rotationally symmetric and
decay fast from a central peak indicating no long-ranged spatial structure. 
However, the ringlike structure of $C_{ww}$
and the maximum of $S_{w}(\vec{k})$ at a finite $|\vec{k}|$ identify a typical length scale of velocity correlations
corresponding to the typical size of clusters~(Fig. \ref{pic_lowrho_modmum}).     

If the dynamics is dominated by anti-alignment (large $\mu_-$), we observe the formation of patches with nematic order
where the polar clusters form lanes with neighboring lanes moving into opposite directions leading to global nematic order. This 
anti-parallel movement distinguishes these patters from the active crystals described in 
\cite{menzel_traveling_2013,menzel_active_2014}. $C_{\rho\rho}$ and $S_\rho$ 
show a high degree of spatial structure corresponding to the characteristic distance between clusters and lanes,
respectively.  
Interestingly, the momentum autocorrelation function
$C_{ww}$ also shows crystal like structure corresponding to a hexagonal lattice 
(Fig. \ref{pic_lowrho_highmum}). We emphasize that the autocorrelation 
functions were obtained from time-averaged density and momentum fields. In general, these 
autocorrelation functions and spectra will differ from ones obtained from a single snapshot at a fixed time $t$.

Surprisingly, the state with global nematic order and local polar order also emerges for anti-alignment only ($\mu_+=0$)
as shown in Fig. \ref{pic_lowrho_highmum}. In this case, particles do not 
interact for distances $\xi_r\leq r_{ji} <\xi_a$ and local
polar order is enforced by the strong anti-alignment at larger distance. The formation of the particular dynamic cluster
state with dense polar clusters moving in lanes is the only possibility to ensure a stationary pattern when
anti-alignment interaction is dominant. Nematic lanes were reported previously 
from other active matter systems (see e.g.
\cite{helbing_freezing_2000,wensink_emergent_2012,mccandlish_spontaneous_2012} 
and \cite{menzel_tuned_2015} for a
recent review as well as references therein), however not in absence of 
an (effective) alignment interaction and never with 
additional periodic order within individual lanes.

\subsection{High density case ($\rho_0=150$)}

\begin{figure}[H]
\begin{center}
\includegraphics[width=\textwidth]{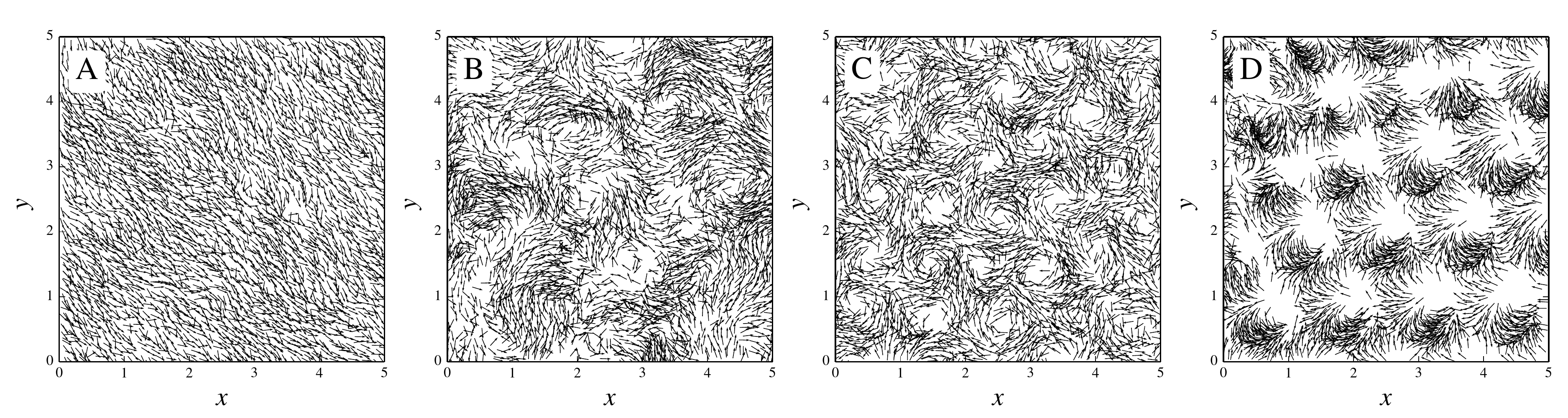}
\caption{Close-up snapshots in the high density regime. From left to right: 
homogeneous polarly ordered state, mesoscale turbulence, rectangular vortex 
array, rotating clusters. }
\label{fig:HD:SNAPSHOTS}
\end{center}
\end{figure}

We discussed simulation results in the high-density regime in \cite{grossmann_vortex_2014}. Here, we review these previous 
observations and go beyond by exploring a wider range of parameters. Snapshots of simulations in the high density regime 
are shown in Fig. \ref{fig:HD:SNAPSHOTS}.

\begin{figure}[p]
 \begin{center}
\includegraphics[width=0.98\textwidth]{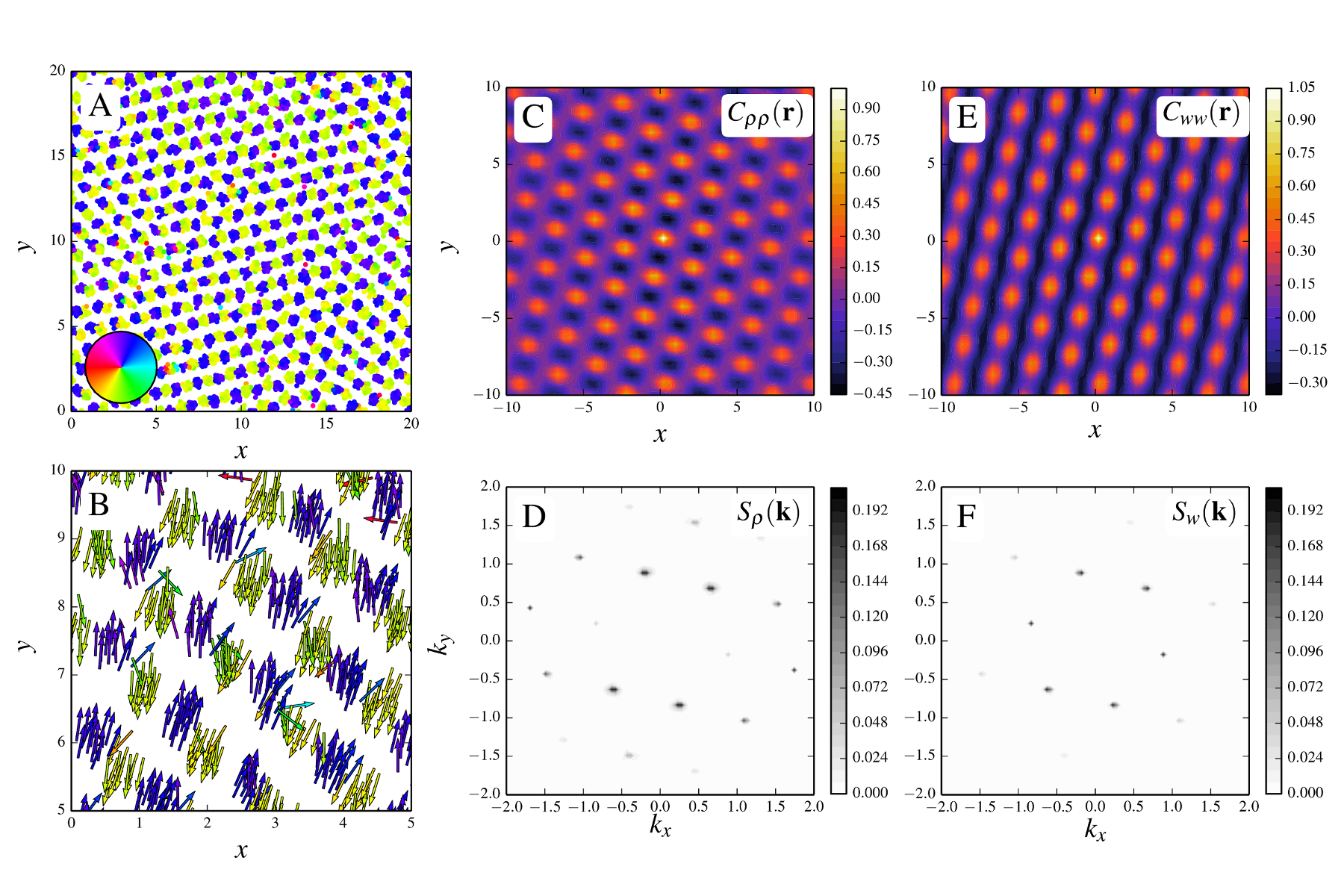}
    \caption{
\textit{Nematic order} {($\Phi_p = 0$ and $\Phi_n = 0.93$)} due to \textit{lanes 
of polar clusters} without alignment
$\mu_+=0$ at strong anti-alignment $\mu_-=4$, density $\rho_0=20$ and noise 
$D_\varphi=1.0$. 
({\bf A}) Global snapshot with particle positions shown by symbols and 
orientation indicated by the color according to
the inset. 
({\bf B}) Zoom into a subregion of size $5\times5$ with particle positions and 
orientations shown as vectors. 
({\bf C}) Time averaged density autocorrelation function $C_{\rho\rho}({\bf r})$ 
and ({\bf D}) its Fourier transform
$S_\rho({\bf k})$. 
({\bf E}) Time averaged momentum field autocorrelation function $C_{ww}({\bf 
r})$ and ({\bf F}) its Fourier transform
$S_v({\bf k})$. 
   \label{pic_lowrho_highmum}
}
 \end{center}
%
%
 \begin{center}
\includegraphics[width=0.98\textwidth]{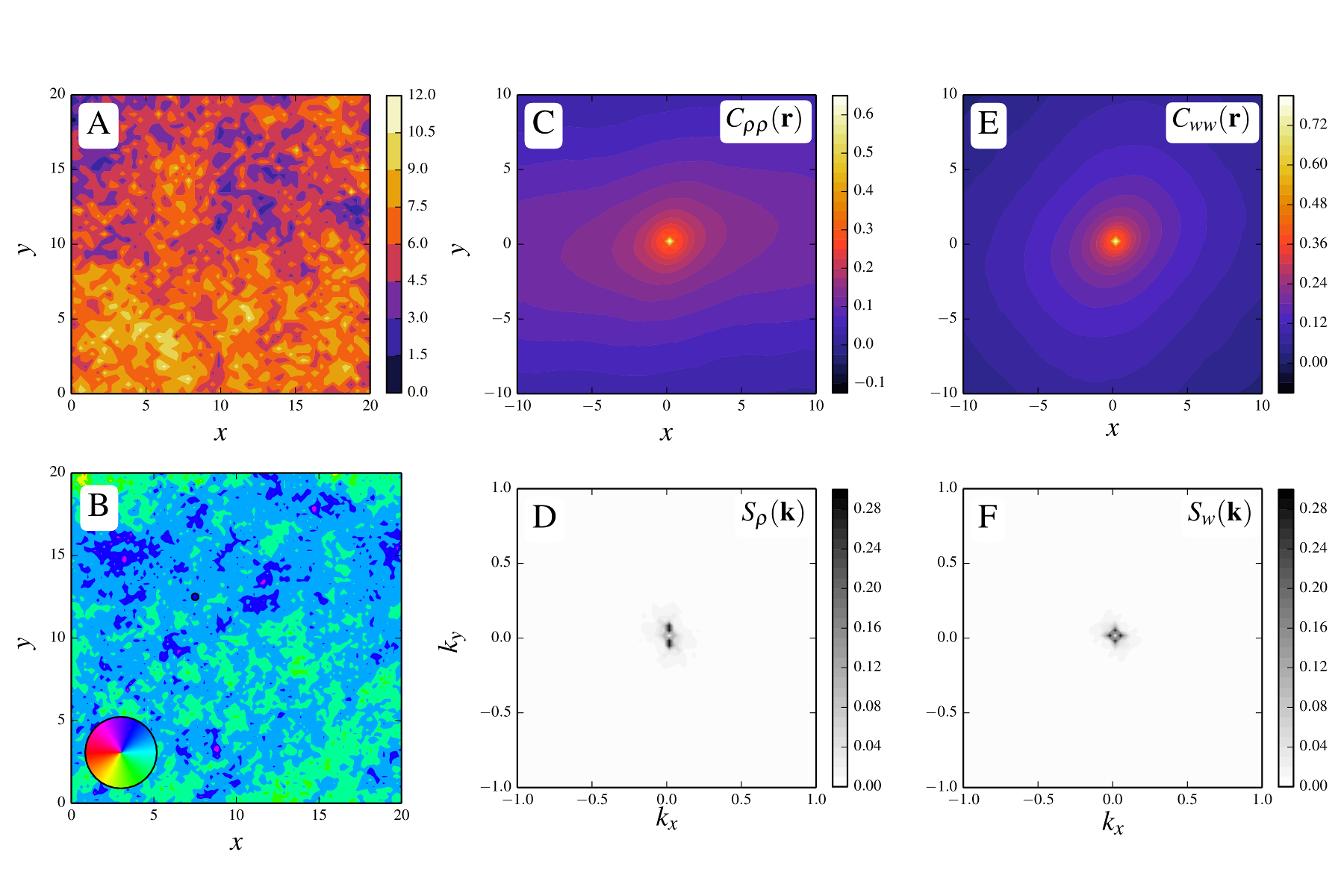}
    \caption{
\textit{Large-scale polar order} {($\Phi_p = 0.92$ and $\Phi_n = 0.73$)} at high 
density $\rho_0=150$ with $\mu_+=1.6$,
$\mu_-=0.0016$ and $D_\varphi=1.0$. 
({\bf A}) Global snapshot of the density field. 
({\bf B}) Global snapshot of the orientation field with polar orientation 
indicated by color according to the inset. 
({\bf C}) Time averaged density autocorrelation function $C_{\rho\rho}({\bf r})$ 
and ({\bf D}) its Fourier transform
$S_\rho({\bf k})$. 
({\bf E}) Time averaged momentum field autocorrelation function $C_{ww}({\bf 
r})$ and ({\bf F}) its Fourier transform
$S_w({\bf k})$. 
   \label{pic_highrho_lowmum}
}
 \end{center}
\end{figure}

\begin{figure}[p]
 \begin{center}
\includegraphics[width=0.98\textwidth]{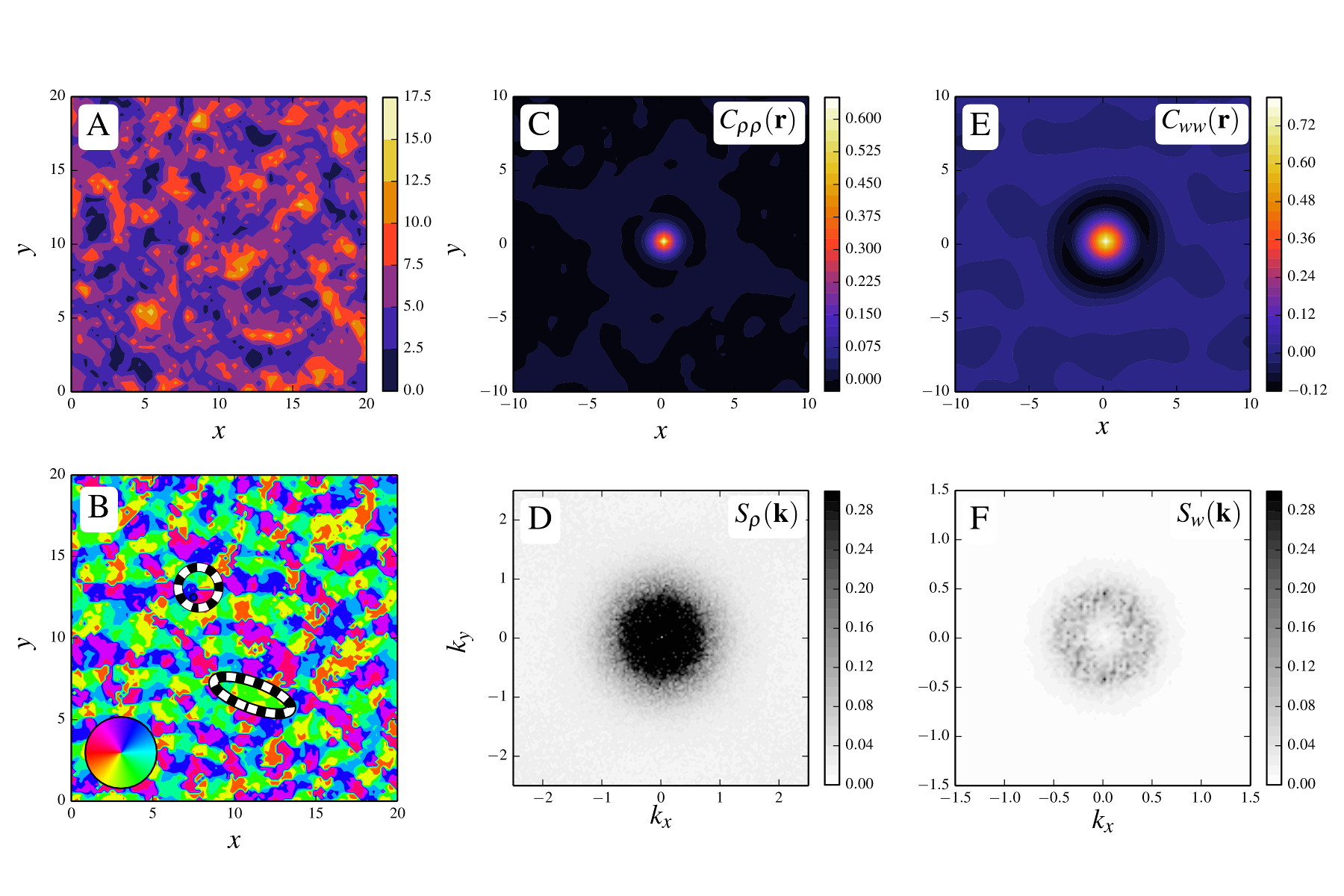}
    \caption{
\textit{Mesoscale turbulence} -- complex spatio-temporal convective flows 
{($\Phi_p = 0.01$ and $\Phi_n = 0.01$)} -- at
high density $\rho_0=150$ with $\mu_+=1.6$, $\mu_-=0.008$ and $D_\varphi=1.0$ 
(see also \cite{grossmann_vortex_2014}). 
({\bf A}) Global snapshot of the density field. 
({\bf B}) Global snapshot of the orientation field with polar orientation 
indicated by color according to the inset. The
black-white elipse indicates a \textit{jet} and the ring a transient vortex. 
({\bf C}) Time averaged density autocorrelation function $C_{\rho\rho}({\bf r})$ 
and ({\bf D}) its Fourier transform
$S_\rho({\bf k})$. 
({\bf E}) Time averaged momentum field autocorrelation function $C_{ww}({\bf 
r})$ and ({\bf F}) its Fourier transform
$S_w({\bf k})$. 
   \label{pic_highrho_mid1mum}
}
 \end{center}
%
 \begin{center}
\includegraphics[width=0.98\textwidth]{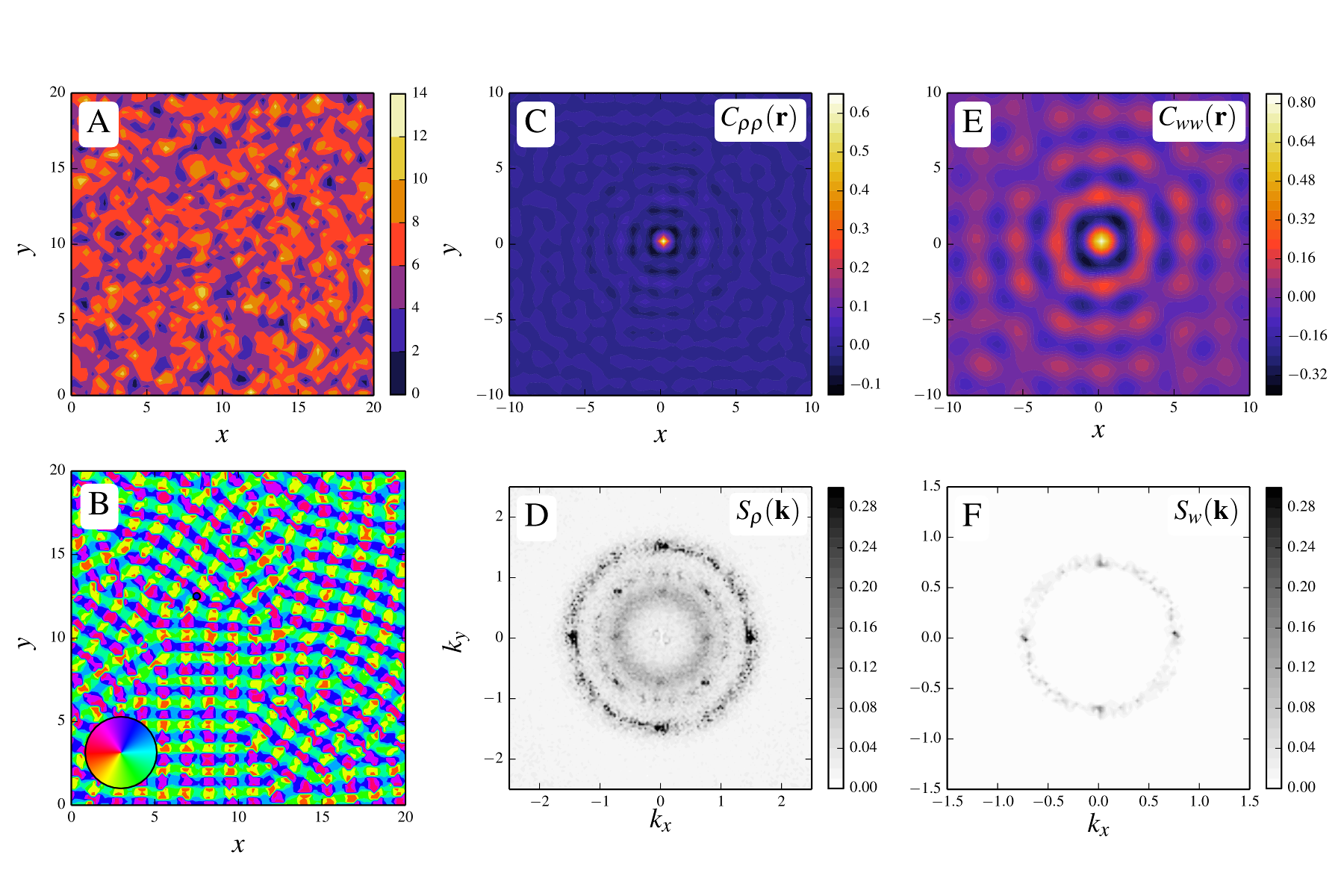}
    \caption{
\textit{Vortex lattice} {($\Phi_p = 0.0$ and $\Phi_n = 0.01$)} with (local) 
rectangular positional order at high
density $\rho_0=150$ with $\mu_+=3.2$,
$\mu_-=0.032$ and
$D_\varphi=1.0$ (see also \cite{grossmann_vortex_2014}). 
({\bf A}) Global snapshot of the density field. 
({\bf B}) Global snapshot of the orientation field with polar orientation 
indicated by color according to the inset. 
({\bf C}) Time averaged density autocorrelation function $C_{\rho\rho}({\bf r})$ 
and ({\bf D}) its Fourier transform
$S_\rho({\bf k})$. 
({\bf E}) Time averaged momentum field autocorrelation function $C_{ww}({\bf 
r})$ and ({\bf F}) its Fourier transform
$S_w({\bf k})$. 
   \label{pic_highrho_mid2mum}
}
 \end{center}
\end{figure}

Similar to the low-density case, we observe the emergence of large scale polar order for vanishing $\mu_-$. However, there are no bands due to
the high density and we observe the Toner-Tu phase (Fig. \ref{pic_highrho_lowmum}). 
An increase in $\mu_-$ beyond a critical value leads to the breakdown of polar order. Just above this transition, we
observe a dynamical regime without any spatial order but with a finite scale
in the momentum field, which is reflected by a single ring of negative correlation about the central maximum of
$C_{ww}$ and the rotationally symmetric $S_w$ with a maximum at finite $|{\bf k}|$. In contrast to the low-density
case, we observe extended convective flows (\textit{jets}) and transient vortices (Fig. \ref{pic_highrho_mid1mum})
instead of diffuse clusters. This is the regime discussed in 
\cite{grossmann_vortex_2014} which
reproduces qualitatively and quantitatively the observations of mesoscale turbulence in dense bacterial suspensions.

\begin{figure}[tb]
 \begin{center}
\includegraphics[width=0.98\textwidth]{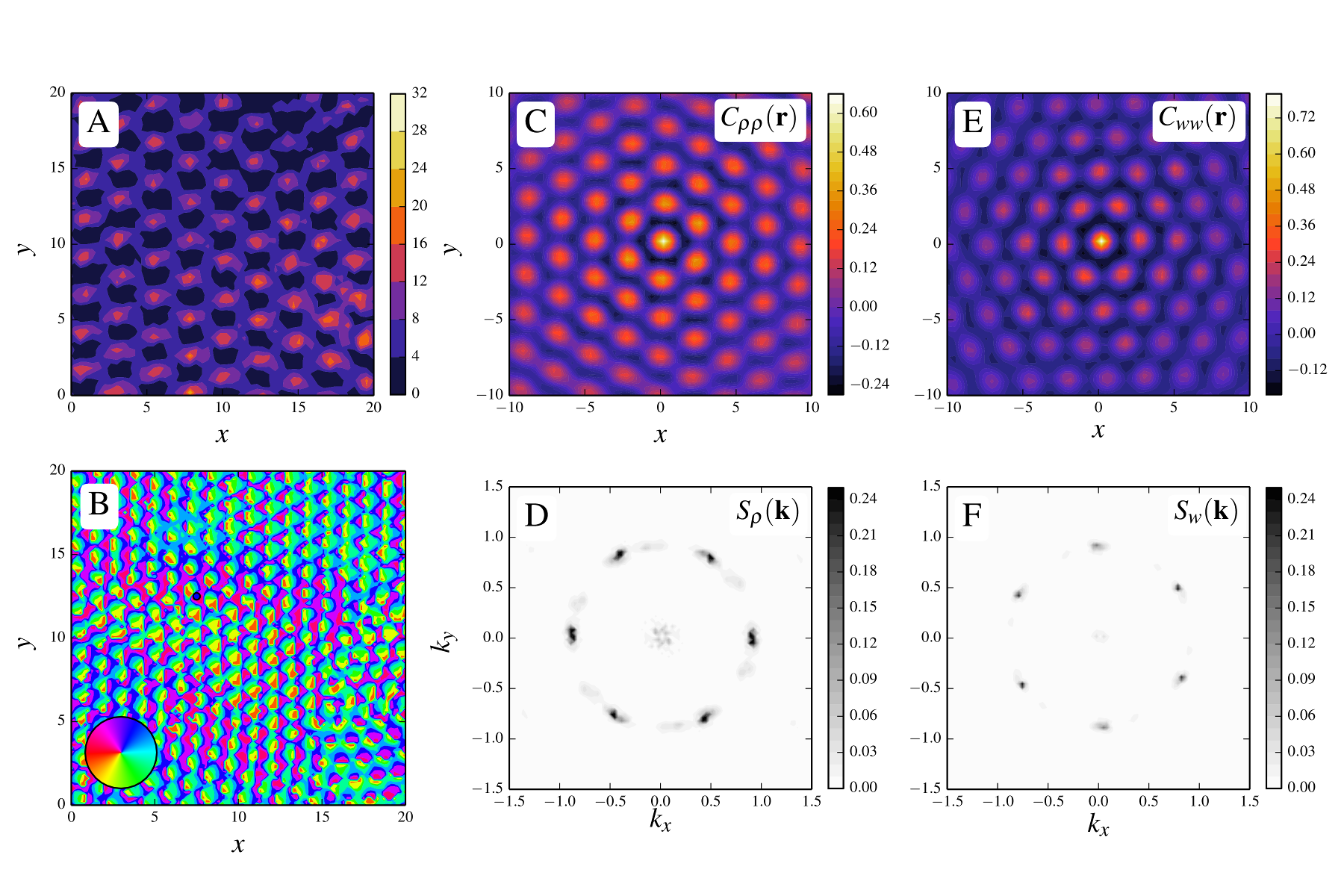}
    \caption{
\textit{Dense rotating clusters} {($\Phi_p = 0.06$ and $\Phi_n = 0.02$)} at high 
density $\rho_0=150$ with $\mu_+=1.6$,
$\mu_-=1.024$ and $D_\varphi=1.0$. The velocity field as well as the density are 
periodically modulated. 
({\bf A}) Global snapshot of the density field. 
({\bf B}) Global snapshot of the orientation field with polar orientation 
indicated by color according to the inset. 
({\bf C}) Time averaged density autocorrelation function $C_{\rho\rho}({\bf r})$ 
and ({\bf D}) its Fourier transform
$S_\rho({\bf k})$. 
({\bf E}) Time averaged momentum field autocorrelation function $C_{ww}({\bf 
r})$ and ({\bf F}) its Fourier transform
$S_w({\bf k})$. 
   \label{pic_highrho_highmum}
}
 \end{center}
\end{figure}

An increase in $\mu_-$ initially leads to the stabilization of the vortices and we observe the emergence of patches of
spatially ordered vortex arrays arranged in a rectangular lattice with rather homogeneous density (Fig.
\ref{pic_highrho_mid2mum}). However, a further increase in $\mu_-$ leads also to growth of density inhomogeneities and
eventually the rectangular vortex pattern transforms into a hexagonal array of rotating clusters (Fig.
\ref{pic_highrho_highmum}). 



\newpage 
\section{Kinetic \& hydrodynamic theory}
\label{sec:theory}

In this section, we turn our attention to the analytical analysis of the microscopic model aiming at a qualitative
understanding of the mechanism leading to pattern formation on macroscopic scales. For that purpose, we systematically
derive kinetic as well as hydrodynamic equations which describe the dynamics of the system in this regime. In
particular, we quantify the quality of the approximations involved in the coarse-graining process and discuss their
range of applicability. Our analysis is based on the study of the one-particle 
distribution function
 \begin{align}
  p(\bi{r},\varphi,t) &= \sum_{i=1}^N \mean{\delta^{(2)}\!\left ( \bi{r} - \bi{r}_i(t) \right) \delta \! \left
(\varphi - \varphi_i(t) \right)} \!, \label{eqn:def:pa} 
 \end{align}
which determines the density of particles at position $\vec{r}$ moving into the direction $\varphi$. Its dynamics cannot
be obtained rigorously in a closed form for an interacting many-particle system. This fact is easily understood: In
order to calculate the force between two particles, one has to know the probability to find a particle at position
$\vec{r}$ given that another particle is located at $\vec{r}'$. Hence, the dynamics of the one-particle distribution
involves the two-particle probability distribution. Here, we overcome this hierarchy problem by assuming that
fluctuations are small (\textit{mean-field assumption}) enabling us to approximate the
two-particle probability distribution by the product of one-particle densities (see appendix \ref{sec:A:MFT} for
details). Hence, our kinetic theory does not account for inter-particle correlations. However, this assumption is
justified in the limit of high particle densities: In this situation, relative density fluctuations are small and,
consequently, every particle interacts with an approximately constant number of neighbors \cite{peruani_mean_2008,
farrell_pattern_2012,
grossmann_self-propelled_2013, grossmann_vortex_2014}. 

Using the mean-field arguments above, we obtain the following nonlinear Fokker-Planck equation for the dynamics of the
one-particle density
 \begin{align}
  \label{eqn:nonl:FPE}
    \frac{\partial p\! \left ( \bi{r},\varphi,t \right )}{\partial t} = &
    - \begin{pmatrix}
       \cos \varphi \\ \sin \varphi
      \end{pmatrix} \! \cdot \! \nabla_{\vec{r}} \, p(\vec{r},\varphi,t)    
%
%
- \frac{\partial}{\partial \varphi} \! \left \{ \! F[p] \, p(\vec{r},\varphi,t) - D_\varphi \frac{\partial
p(\vec{r},\varphi,t)}{\partial \varphi} \right \} \! 
%
 \end{align}
where the nonlinearity enters through the force 
 \begin{align}
 \label{eqn:MF:force:FPE}
  F[p] = \int \! \rmd^2 r' \! \! \int_0^{2 \pi} \! \rmd \varphi' \left \{ \kappa \! \left ( \abs{\bi{r}'} \right )
\sin \! \left [ \varphi - \arg(\vec{r}') \right ] + \mu \! \left(\abs{\bi{r}'}
\right ) \sin \! \left [ \varphi' - \varphi \right] \right \}  \! p \! \left (\bi{r}\!+\!\bi{r}',\varphi',t \right ) \!
.
 \end{align}
The terms in Eq. \eqref{eqn:nonl:FPE} arise from the following microscopic rules: The first term describes
convection due to active motion at constant speed, the second term reflects the binary, nonlocal
interaction (in mean-field approximation) and the
third term accounts for angular fluctuations. This nonlinear Fokker-Planck equation is solved by the spatially
homogeneous, rotationally invariant state  
 \begin{align}
  \label{eqn:stat:sol:15}
  p_0 = \frac{\rho_0}{2 \pi}
 \end{align}
which describes a disordered configuration without orientational order and a spatially homogeneous particle
density $\rho_0$. In the following paragraphs, we study the
linear stability of this fixed point considering the full dynamics \eqref{eqn:nonl:FPE} and compare the results to the
predictions of the linear stability analysis using a reduced set of equations (\textit{hydrodynamic theory}). This
comparison allows a critical assessment of the approximations used in the derivation of the hydrodynamic theory. 

\subsection{One-particle Fokker-Planck equation in Fourier space} 
\label{sec:FPE:FS}

In order to study the nonlinear Fokker-Planck equation \eqref{eqn:nonl:FPE} analytically, it is convenient to work in
Fourier space with respect to the angular variable $\varphi$. 
%
%
Here, we use the following convention for the Fourier transformation: 
 \begin{align}
   \label{eqn:def:FT:angle}
   \fourier{f}{n}(\bi{r},t) = \int_0^{2 \pi} \rmd \varphi \, e^{i n \varphi} \, p(\bi{r},\varphi,t) \, .
 \end{align}
Due to the nonlocality of the Fokker-Planck equation~--~the
dynamics of $p(\bi{r},\varphi,t)$ depends on the integral over $p(\bi{r},\varphi,t)$ itself~--~the Fourier
transformation is somewhat involved. 
It is helpful to express the force $F[p]$, Eq. \eqref{eqn:MF:force:FPE}, in terms of Fourier modes first.  
For this purpose, we assume that the one-particle density $p(\vec{r}+\vec{r}',\varphi,t)$ can be expanded in a Taylor
series\footnote{Furthermore, we assume the global convergence of the Taylor series. } in the spatial coordinate
$\vec{r}'$ around $\vec{r}' = 0$. 
%
%
In other words, we transform the integral operator in a series of differential operators. We keep all orders in the
Taylor expansion and obtain
 \begin{align}
  \label{eqn:MF:Fourier05}
  F = \frac{\pi}{i} \left [ e^{-i \varphi} \left ( \hat{\mu}_{\Delta} \fourier{f}{1}(\vec{r},t) - \mathcal{K}_{\Delta}
\nabcp \fourier{f}{0}(\vec{r},t) \right ) - \mbox{c.c} \right ] \! .
 \end{align}
In \eqref{eqn:MF:Fourier05}, we introduced the complex derivative $\nabcp = \frac{1}{2} \left ( \partial_x + i
\partial_y \right )$ as well as the operators 
  \begin{align}
 \label{eqn:def:hat:mu:D}
   \hat{\mu}_{\Delta} &= \int_0^{\infty} \rmd r \, r \mu(r) J_0 \! \left ( r \sqrt{-\Delta} \right ) 
    = \sum_{n=0}^{\infty} \left [ \int_0^{\infty} \rmd r \, r \, \mu(r) \, \frac{r^{2n}}{4^n (n!)^2} \right ] \!
\Delta^{n} ,
  \end{align}
and 
 \begin{align}
 \label{eqn:def:hat:K:D}
  \mathcal{K}_{\Delta} = 2 \int_0^{\infty} \rmd r \, r \kappa(r) \frac{J_1 \! \left ( r \sqrt{-\Delta}\right
)}{\sqrt{-\Delta}} = \sum_{n=0}^\infty \left [ \int_0^\infty \rmd r \, r \, \kappa(r) \, \frac{r^{2n+1}}{4^n n! (n+1)!}
 \right ] \! \Delta^{n} 
 \end{align}
which stem from the Taylor expansion. We denote the Laplace operator by $\Delta$. The first term in
\eqref{eqn:def:hat:mu:D} and \eqref{eqn:def:hat:K:D} is written in a formal way using Bessel functions of the first kind
$J_\nu(x)$, whereas the second follows immediately from the series expansion of the Bessel functions. These series
reflect the nonlocality of the interaction. 
Interestingly, the alignment interaction couples to the first Fourier mode only (proportional to the polar order
parameter) and the repulsive interaction couples to the zeroth mode (particle density).

Having rewritten the force $F[p]$ as it is done in Eq. \eqref{eqn:MF:Fourier05}, the dynamics of the angular Fourier
coefficients is directly obtained by taking the temporal derivative of \eqref{eqn:def:FT:angle} and inserting the
nonlinear Fokker-Planck equation for the one-particle density on the right hand side: 
 \begin{align}
   \label{eqn:dyn:FT:angle}
      \frac{\partial \fourier{f}{n}}{\partial t} = &- \left ( \nabcp \fourier{f}{n-1} + \nabcp^* \fourier{f}{n+1}
\right) + n \pi \! \left [ \fourier{f}{n-1} \, \hat{\mu}_{\Delta} \, \fourier{f}{1} - \fourier{f}{n+1} \,
\hat{\mu}_{\Delta} \, \fourier{f}{-1} \right ] \nonumber \\
      & - n \pi \left [ \fourier{f}{n-1} \, \mathcal{K}_{\Delta} \nabcp \fourier{f}{0} - \fourier{f}{n+1} \,
\mathcal{K}_{\Delta} \nabcp^{*} \fourier{f}{0}  \right ] - n^2 D_{\varphi} \fourier{f}{n}.
 \end{align}

The dynamics of the $n$th Fourier coefficient \eqref{eqn:dyn:FT:angle} is in general coupled to others meaning that
the nonlinear, nonlocal Fokker-Planck equation is equivalent to an infinite hierarchy of equations in Fourier space.
This hierarchy is, however, insightful because it allows the investigation of the consequences of approximations,
such as the  
\begin{itemize}
 \item negligence of Fourier coefficients $\fourier{f}{n}$ for $\abs{n} > n_c$, where $n_c > 0$, 
 \item negligence of derivatives of a particular order,
\end{itemize}
which are studied within linear stability analysis of the spatially homogeneous state in the following paragraphs.

\subsection{Linear stability analysis within kinetic theory}
\label{sec:lin:stab:ana:fullFP}

Before we discuss the consequences of several approximations, we illustrate the predictions of the full kinetic theory
\eqref{eqn:dyn:FT:angle} in this section. In particular, we study the linear stability of the disordered state by
linearizing the kinetic equations for small perturbations. In Fourier space, the dynamics of the
perturbations $\delta \fourier{f}{n}(\bi{k},t)$ is determined by a linear system of coupled differential equations
 \begin{align}
  \frac{\rmd \delta \fourier{f}{n}(\bi{k},t)}{\rmd t} = \sum_{l=-\infty}^\infty \tilde{M}_{nl} \! \left ( \abs{\bi{k}}
\right ) \delta \fourier{f}{l}(\bi{k},t)
 \end{align}
with the coefficients
 \begin{align}
  \label{eqn:matrix:elments:Mnl}
%
 \tilde{M}_{nl} \! \left ( \abs{\bi{k}} \right ) =& \frac{i}{2} \left [ (k_x + i k_y) \delta_{n-1,l} + (k_x - i k_y)
\delta_{n+1,l} \right ] + n \pi \rho_0 \hat{\mu}_k \left ( \delta_{n,1} - \delta_{n,-1}\right) \delta_{n,l}  \nonumber
\\
    & - n^2 D_\varphi \delta_{n,l} + \frac{i n \pi \rho_0}{2} \mathcal{K}_k \left [ (k_x + i k_y) \delta_{n,1} - (k_x -
i k_y) \delta_{n,-1} \right ] \delta_{l,0} \, .
 \end{align}
In \eqref{eqn:matrix:elments:Mnl}, the Fourier representation (spectra) of the nonlocal operators $\hat{\mu}_\Delta$ and
$\mathcal{K}_\Delta $ were introduced: 
  \begin{align}
    \hat{\mu}_{k}   &= \int_0^{\infty} \rmd r \, r \mu(r) J_0 \! \left ( \abs{\bi{k}} \! r \right ) \!,
\label{eqn:op:mu_K} \quad 
 \mathcal{K}_{k} = 2 \int_0^{\infty} \rmd r \, r \kappa(r) \frac{J_1 \! \left ( \abs{\bi{k}} \! r \right
)}{\abs{\bi{k}}} \, .
\end{align}
The eigenvalues of the matrix $\tilde{M}$ as a function of the wavevector $\vec{k}$~--~also referred to as
dispersion relation~--~determine the stability of the disordered state. If the real part of an eigenvalue is positive,
the disordered state is unstable.  %
Since the disordered state is invariant under spatial rotations, the dispersion relation does depend on the wavenumber
$\abs{\vec{k}}$ only. 
%
We denote the dispersion relation by $\sigma\! \left ( \abs{\bi{k}} \right )$ and its real part by $\Re[\sigma\! \left (
\abs{\bi{k}} \right )]$. Once $\sigma\! \left ( \abs{\bi{k}} \right )$ has been calculated, we can distinguish different
scenarios: 
 \begin{enumerate}
  \item[(i)] If $\Re[\sigma\! \left (
\abs{\bi{k}} \right )]$ is positive at its maximum located at $\abs{\bi{k}} = 0$, the
spatially homogeneous system is unstable.  
  \item[(ii)] If $\Re[\sigma\! \left (
\abs{\bi{k}} \right )]$ is positive at its maximum located at a finite value of
$\abs{\bi{k}}$, a pattern with a finite characteristic length scale is expected to emerge (Turing instability). 
  \item[(iii)] If $\Re[\sigma\! \left (
\abs{\bi{k}} \right )]$ is non-positive, the disordered state is stable.  
 \end{enumerate}
Information about case (i) is obtained from the study of the spatially homogeneous system by setting
$\abs{\vec{k}} = 0$. The matrix \eqref{eqn:matrix:elments:Mnl} reduces to a diagonal matrix in this case. Hence, the
eigenvalues are trivially obtained. The analysis reveals that only the first Fourier coefficient (related to the polar
order parameter) can become unstable. This instability is due to the alignment interaction which does only couple to the
first Fourier coefficient as can be seen from Eq. \eqref{eqn:MF:Fourier05}. 
The parameter region where this instability occurs~--~signaling the emergence of local polar order~--~is determined by 
 \begin{align}
  \label{eqn:order:dis:trans:line}
  \pi \rho_0 \left ( \int_0^{\infty} \rmd r \, r \mu(r) \right) - D_\varphi  > 0. 
 \end{align}
Equation \eqref{eqn:order:dis:trans:line} is a generalization of the expression derived in
\cite{peruani_mean_2008} for an arbitrary distance dependent velocity-alignment strength. Consistently, spatially homogeneous, polarly ordered states
cannot emerge if the interaction is dominated by anti-alignment, i.e. if the condition $\int_0^{\infty} \rmd r \, r
\mu(r) < 0$
is fulfilled. 

\begin{figure}[p]
 \begin{center}
    \includegraphics[width=0.95\textwidth]{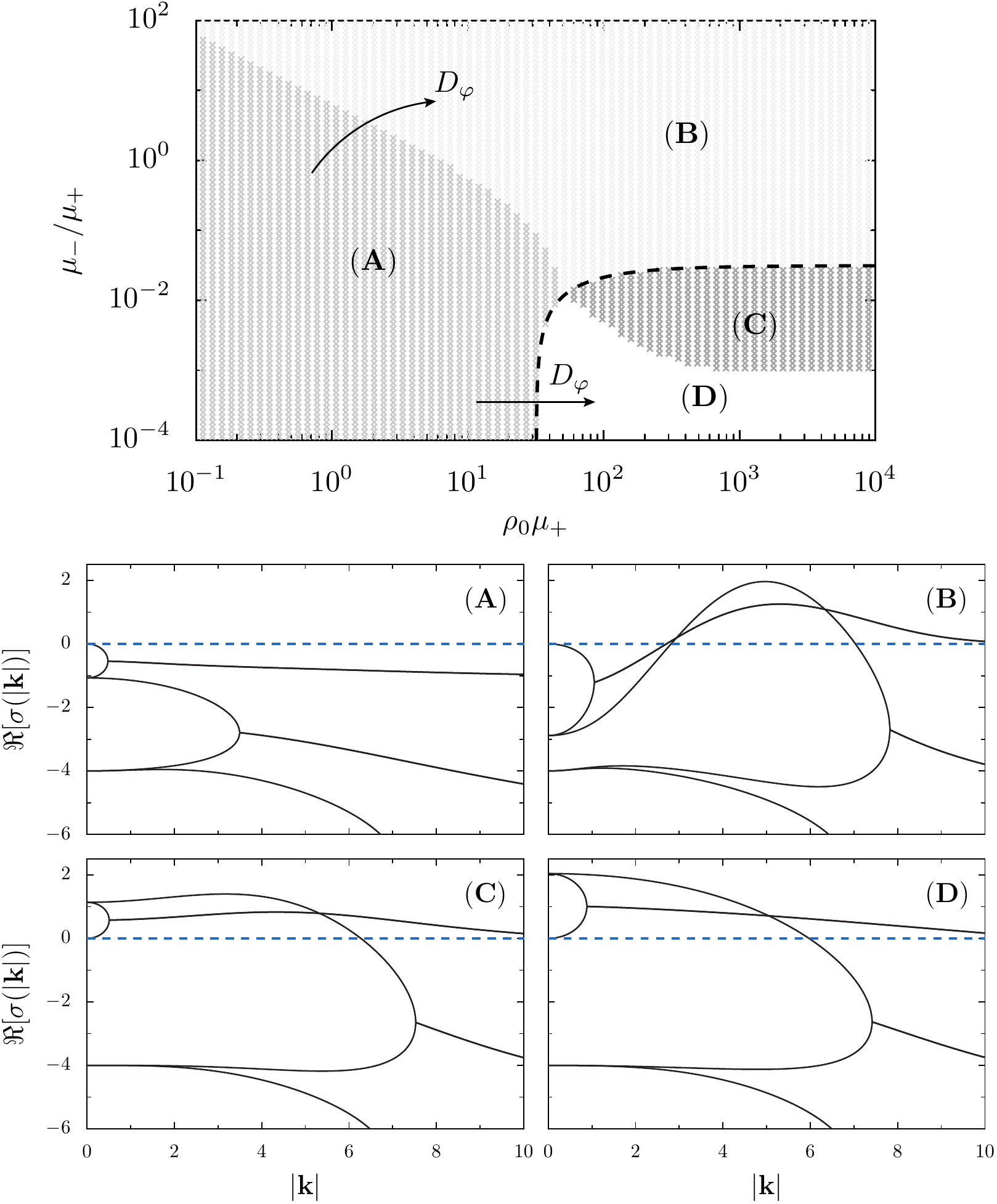}
   \caption{Results of the numerical linear stability analysis of the disordered state $p_0$, Eq. \eqref{eqn:stat:sol:15}, within 
kinetic
theory. \textbf{Top}: The ($\rho_0\mu_+$)-($\mu_-/\mu_+$)-plane of the parameter space. \textbf{Bottom}:
Exemplary dispersion relations $\Re[\sigma(\abs{\bi{k}})]$ obtained numerically from \eqref{eqn:matrix:elments:Mnl}. If
the dispersion relation is positive, the disordered state undergoes an instability.  
In parameter region $(A)$, the disordered state is stable. We find Turing
instabilities in the parameter range $(B)$. In $(C)$, the dispersion relation possesses a maximum at finite $\abs{\vec{k}}$ and, 
additionally,
the homogeneous system ($\abs{\vec{k}} = 0$) is unstable. In region $(D)$, the spatially homogeneous system is
unstable.   
%
%
The boundary $\sigma(\abs{\vec{k}}=0) = 0$, determined by \eqref{eqn:order:dis:trans:line}, is represented by a dashed
line. Arrows in top panel indicate the shift of critical lines when the noise intensity $D_\varphi$ is
increased. Parameters of dispersion relations: $\rho_0 \mu_+ = 1$, $\mu_-/\mu_+ = 10^{-1}$ in (A);
$\rho_0 \mu_+ = 10^2$, $\mu_-/\mu_+ = 5 \cdot 10^{-2}$ in (B); $\rho_0 \mu_+ = 10^2$, $\mu_-/\mu_+ = 10^{-2}$ in (C);
$\rho_0 \mu_+ = 10^2$, $\mu_-/\mu_+ = 10^{-3}$ in (D); $D_\varphi = 1.$, $\xi_a=0.2$, $\xi_r = 0.1$, $\kappa_0 \rho_0 =
1.5 \cdot 10^3$. }
   \label{pic:stab:ana:FP}
 \end{center}
\end{figure}

The full dispersion relation cannot be obtained analytically. In order to study the
linear stability of the disordered state with respect to spatially inhomogeneous perturbations, we calculated the
dispersion relation numerically. Since the matrix $\tilde{M}$ is infinite dimensional, it
must be
truncated first. However, we can exploit that the Fourier modes $\fourier{f}{n}$ are damped proportional to $n^2$ due to
angular
fluctuations. Consequently, high order Fourier modes are strongly damped. Accordingly, we observe that the largest
eigenvalues of the matrix \eqref{eqn:matrix:elments:Mnl} do not depend on the truncation used if a reasonably high
number of Fourier modes is considered. Here, we reduce $\tilde{M}$ to a
$(101\times101)$-matrix. The results of the linear stability analysis are shown in Fig. \ref{pic:stab:ana:FP}. 

The linear stability analysis reveals the following regimes: If the alignment-strength and the density are small, the
disordered state is linearly stable (parameter region (A) in Fig. \ref{pic:stab:ana:FP}). An instability at finite
wavelengths (\textit{Turing instability}) appears if the anti-alignment strength is high (domain (B)). This
instability signals the transition towards an ordered state which possesses a characteristic length scale (cf.
numerical simulation in section \ref{sec:sim}). The disordered state is unstable with respect to spatially homogeneous
perturbations if the alignment strength is high and anti-alignment is negligible (domain (D)).
%
%
In between region (B) and (D), an intermediate regime (C) exists where the dispersion relation is positive at
$\abs{\vec{k}} = 0$ but possesses a maximum at a nonzero wavenumber. 
%

The occurrence of a Turing instability in our model is due to the
concurrence of short-range alignment and anti-alignment interaction. This instability can only be
found if the anti-alignment interaction is sufficiently strong. Consequently, this instability mechanism is not present
in models with purely aligning interactions. 

From the linear stability analysis, we cannot rigorously infer the structure of the emergent pattern which crucially
depends on nonlinear terms. However, we may expect the emergence of collective states which possess a characteristic
length scale in region (B) and (C). Indeed, we observe these state such as clusters of characteristic sizes, mesoscale
turbulence and vortex arrays in Langevin simulations of the particle dynamics (cf. section \ref{sec:sim}, in particular Fig. 
\ref{fig:PD_Sim}). 
%

\subsection{Hydrodynamic theory}
\label{sec:hydrodynamic:limits}

The linear stability analysis of the disordered state within kinetic theory revealed the existence of several
instability mechanism. However, the dispersion relations could only be obtained numerically. In this section, we
discuss the possibility to describe the instabilities of the disordered state analytically by a reduced set of
linearized hydrodynamic equations aiming at a qualitative understanding of the macroscopic pattern formation process. 

The basic assumption of the hydrodynamic theory is that only those observables in the hierarchy
\eqref{eqn:dyn:FT:angle} are important whose large scale dynamics is slow either due to the fact that they describe a
conserved quantity or they describe broken continuous symmetries. For the identification of the relevant degrees of
freedom in our model, we exploit that (i) the \textit{particle number is conserved} and (ii) the microscopic interaction
favors \textit{local polar order} due to velocity-alignment. Hence, the relevant order parameters are the particle
density 
 \begin{align}
   \rho(\vec{r},t) = \fourier{f}{0}(\vec{r},t) = \int_0^{2\pi} \rmd \varphi \, p(\vec{r},\varphi,t)
 \end{align}
and the momentum field  
\begin{subequations} 
 \begin{align}
 \vec{w} (\bi{r},t) & = \int_0^{2\pi} \rmd \varphi \, p(\vec{r},\varphi,t) \begin{pmatrix}
                                                                            \cos \varphi \\ \sin \varphi
                                                                           \end{pmatrix}
 = \begin{pmatrix}
    \Re[\fourier{f}{1}(\bi{r},t)] \\ \Im[\fourier{f}{1}(\bi{r},t)]
   \end{pmatrix} \! ,
%
%
 \end{align}
\end{subequations}
which is proportional to the local flow field $\vec{v}(\vec{r},t) = \vec{w}(\vec{r},t)/\rho(\vec{r},t)$. Please note
that we consider the order parameters locally in space. The Langevin simulations of section \ref{sec:sim} revealed
that all macroscopic patterns display local polar order. This assumption does not exclude the description of
nematic states such as polar clusters moving in opposite directions (cf. Fig. \ref{pic_lowrho_highmum}) which possess
local polar and global nematic order. 

In general, the dynamics of the density and the momentum field is coupled to higher order Fourier modes. In particular,
the dynamics of the polar and nematic order parameter is not independent. However, we argued above that the dynamics of
the relevant order parameters is slow on large scales. The assumption of fast relaxation of higher order Fourier modes
enables us to truncate the hierarchy \eqref{eqn:dyn:FT:angle} at low order. In particular, we express the nematic order parameter 
as a function
of the polar order parameter and the density by adiabatic elimination, and further neglect all higher order Fourier
coefficients following the
same reasoning. Here, we focus on linear terms since we are
interested in comparing the predictions of the linear stability analysis of the disordered state. We obtain the
following set of linearized hydrodynamic equations: 
\begin{subequations} 
\label{eqn:hydro:theor:real}
 \begin{align}
   \frac{\partial \rho}{\partial t}   = &- \nabla \cdot \bi{w} , \label{ref:contEQ} \\
   \frac{\partial \bi{w}}{\partial t} \approx & - \left ( \frac{1}{2} + \frac{\pi \rho_0 \kappa_0 \xi_r^3}{6} \right  )
\!
\nabla \rho - D_\varphi \bi{w} + \pi \rho_0 \hat{\mu}_{\Delta}
\bi{w} + \frac{\Delta \bi{w}}{16 D_\varphi} \, . \label{eqn:hydro:theor:realB}
 \end{align}
\end{subequations}
The continuity equation \eqref{ref:contEQ} reflects the conservation of the number of particles. The equation for the
momentum field contains several physical effects. The first term which is proportional to the gradient of the density
describes a macroscopic particle flow compensating density gradients. The prefactor in brackets is proportional to the
pressure which contains two contributions: (i) the constant term is due to the active motion; (ii) the second
term is proportional to the repulsive force which effectively increases the pressure, i.e. it reduces the
compressibility. The second term in
\eqref{eqn:hydro:theor:realB} accounts for the effect of noise counteracting the emergence of polar order. The nonlocal
alignment interaction is described by the operator $\hat{\mu}_\Delta$ defined via a series of Laplacian
operators \eqref{eqn:def:hat:mu:D}. The last term represents the reduction of the effective viscosity due to
angular noise. It is the only term which is influenced by the truncation scheme used to derive
\eqref{eqn:hydro:theor:real}. 

\subsubsection*{Linear stability analysis within the hydrodynamic theory}

The truncation of the infinite hierarchy of Fourier modes to the simplified system of hydrodynamic equations is a
considerable simplification. In particular, it allows the analytical calculation of the dispersion relations which can
be directly obtained by Fourier transformation of the hydrodynamic equations \eqref{eqn:hydro:theor:real} and subsequent matrix 
diagonalization: 
 \begin{subequations}
 \label{eqn:eigval:hydro}
   \begin{align}
     \sigma_1 \! \left( \abs{\bi{k}} \right )     &= \pi \rho_0 \hat{\mu}_k - D_\varphi - \frac{\abs{\bi{k}}^2}{16
D_\varphi} , \\
     \sigma_{2,3} \! \left( \abs{\bi{k}} \right ) &= \frac{1}{2} \left [ \sigma_1\! \left( \abs{\bi{k}} \right ) \pm
\sqrt{ \left [\sigma_1\! \left( \abs{\bi{k}} \right ) \right ]^2 - 2 \left ( 1 + \frac{\pi \rho_0 \kappa_0
\xi_r^3}{3} \right  ) \! \abs{\bi{k}}^2} \,\, \right ] \! .
   \end{align}
 \end{subequations}
We address the question in which parameter range this approximation is applicable and whether these approximated
dispersion relations can be used to describe the destabilization of the disordered state. 

\begin{figure}[tb]
 \begin{center}
    \includegraphics[width=\textwidth]{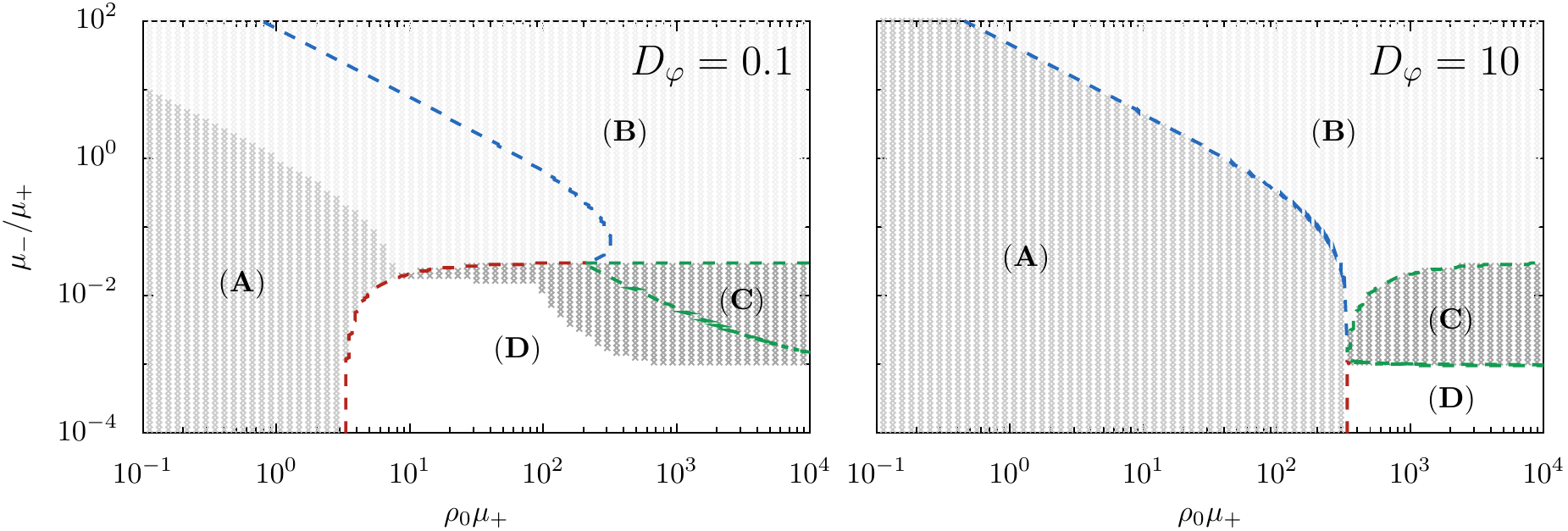}
   \caption{Comparison of the results of the linear stability analysis of the spatially homogeneous state using the
Fokker-Planck equation \eqref{eqn:nonl:FPE} and the hydrodynamic approximation \eqref{eqn:hydro:theor:real}. The results
corresponding to the kinetic theory are shown in the same way as in Fig. \ref{pic:stab:ana:FP}. Predictions
of the hydrodynamic theory are shown by lines. In the high noise limit (right), the hydrodynamic approximation is exact,
whereas deviations are found for low noise values (left). Parameters: $D_\varphi=0.1$ (left), $D_\varphi = 10$ (right).
Other parameters: see Fig. \ref{pic:stab:ana:FP}. }
   \label{pic:stab:ana:hydro:full}
 \end{center}
\end{figure}
The results of the parameter study are illustrated in Fig. \ref{pic:stab:ana:hydro:full}. The analysis was performed
for high and low noise values $D_\varphi$. For high noise amplitudes, we obtain perfect agreement of the predictions of
the hydrodynamic equations \eqref{eqn:hydro:theor:real} with the results of the linear stability analysis which was
based on the one-particle Fokker-Planck equation \eqref{eqn:nonl:FPE}. However, the
predictions of the hydrodynamic theory deviate from the kinetic theory in the limit of low noise values. These
deviations are due to the adiabatic elimination of the second Fourier mode and the negligence of all higher Fourier
coefficients. These deviations could have been expected: The assumption that high order Fourier modes relax fast is
only justifiable for high noise values, since $D_\varphi$ determines the relaxation rate of the Fourier
coefficients as can be seen from \eqref{eqn:dyn:FT:angle}. 

Consequently, we conclude that it is possible to describe the linear dynamics of the system by the hydrodynamic
equations \eqref{eqn:hydro:theor:real} as long as the noise strength is high, i.e. a considerable timescale separation
between the lowest and higher Fourier coefficients exists. 

\subsubsection*{Approximating the nonlocal alignment interaction}

As already mentioned in section \ref{sec:FPE:FS}, one can possibly approximate the infinite series $\hat{\mu}_{\Delta}$
of Laplacian operators by a truncated polynomial: 
 \begin{align}
   \label{eqn:trunc:muD}
   \hat{\mu}_{\Delta} \approx \hat{\mu}_{\Delta,N_c} = \sum_{n=0}^{N_c} \left [ \int_0^{\infty} \rmd r \, r \mu(r) \,
\frac{r^{2n}}{4^n (n!)^2} \right ] \! \Delta^{n} . 
 \end{align}
Naturally, this raises the question how to choose $N_c$ (order of truncation) and how reliable the
corresponding approximations are. The stability of the dynamics requires that
 \begin{align}
   \label{eqn:stab:requirement:Nc}
   (-1)^{N_c} \left [ \int_0^{\infty} \rmd r \, \mu(r) \, r^{2 N_c + 1}  \right ] < 0
 \end{align}
holds. As a rule of thumb, \eqref{eqn:stab:requirement:Nc} reflects that $N_c$ must be odd if alignment interaction
dominates and $N_c$ must be even if anti-alignment is predominant. 

What is the nature of an approximation like \eqref{eqn:trunc:muD}? The operator $\hat{\mu}_{\Delta}$ describing the
nonlocal (anti-)alignment interaction is linear and, thus, it possesses a characteristic spectrum $\hat{\mu}_k$
introduced in Eq. \eqref{eqn:op:mu_K}. Approximating the operator by a series of Laplacian operators,
cf. \eqref{eqn:trunc:muD}, is equivalent to approximate the spectrum by a Taylor series in powers of the
wavenumber $\abs{\vec{k}}$ around $\abs{\vec{k}} = 0$ restricting the applicability of this approximation to small
wavenumbers (large length scales). The stability
requirement \eqref{eqn:stab:requirement:Nc} ensures that the spectrum at large wavenumbers is negative. Otherwise,
fluctuations on small length-scales would be amplified without any bound.  

On phenomenological grounds, there are two conceivable types of spectra: 
\begin{enumerate}
 \item[(i)] the spectrum possesses a maximum at $\abs{\vec{k}} = 0$ and decreases monotonically;
 \item[(ii)] the spectrum possesses a maximum at a finite wavenumber. 
\end{enumerate}
The former case was studied in the context of the Vicsek model (pure alignment interaction \cite{vicsek_novel_1995})
in the corresponding field theory by Toner and Tu \cite{toner_LRO_1995}. The latter was recently proposed by Dunkel et
al. \cite{dunkel_minimal_2013} in order to describe mesoscale turbulence. 
%
%
Using the expansion of the operator
$\hat{\mu}_\Delta$, we can map our linearized hydrodynamic theory to these phenomenological equations
 \begin{subequations}
 \label{eqn:lin:hydrodyn:eqn:1545}
  \begin{align}
    \frac{\partial \rho}{\partial t}   =       &- \nabla \cdot \bi{w} , \\
    \frac{\partial \bi{w}}{\partial t} \approx & - \left ( \frac{1}{2} + \frac{\pi \rho_0 \kappa_0 \xi_r^3}{6} \right  )
\! \nabla \rho + \alpha \hspace{0.03cm} \bi{w}
+ \Gamma_0 \Delta \bi{w} + \Gamma_2 \Delta^{2} \bi{w} ,
 \end{align}
 \end{subequations}
 and derive the dependency of transport coefficients on microscopic model parameters 
\begin{subequations}
 \begin{align}
    \alpha   &= \pi \rho_0 \left ( \int_0^{\infty} \rmd r \, r \mu(r) \right ) - D_\varphi , \\
    \Gamma_0 &= \frac{\pi \rho_0}{4} \left (\int_0^{\infty} \rmd r \, r^3 \mu(r) \right ) + \frac{1}{16 D_\varphi} \, ,
\\
    \Gamma_2 &= \frac{\pi \rho_0}{64} \int_0^{\infty} \rmd r \, r^5 \mu(r) .
 \end{align}
 \end{subequations}
The linearized hydrodynamic equations can describe a Turing instability of the disordered state if the
effective viscosity $\Gamma_0$ is negative \cite{dunkel_minimal_2013}, i.e. only if the anti-alignment interaction is
sufficiently strong. 
 
 \begin{figure}[tb]
 \begin{center}
    \includegraphics[width=\textwidth]{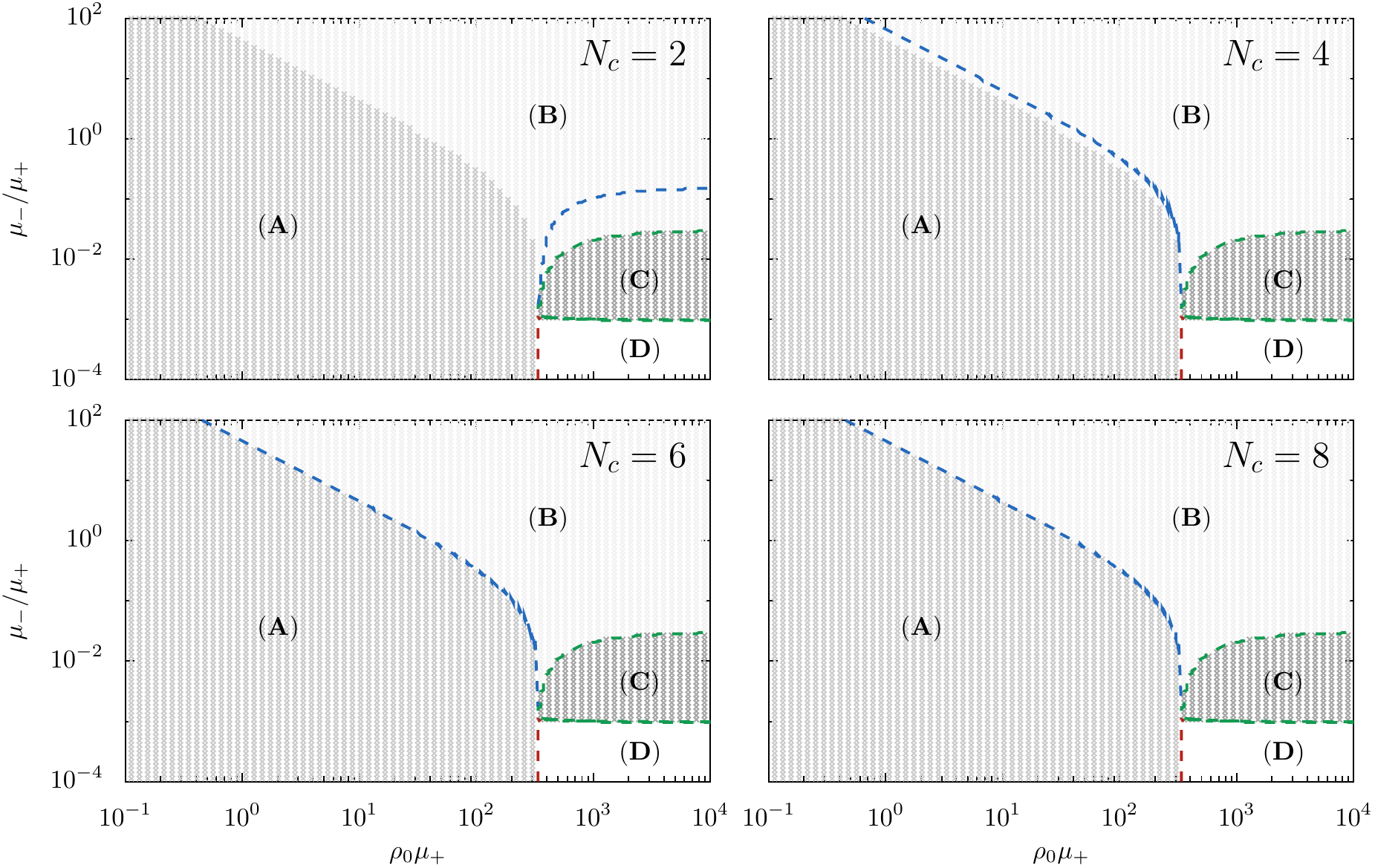}
   \caption{Domains of stability and instability of the disordered state. Predictions of the kinetic theory
\eqref{eqn:nonl:FPE} are depicted as gray shaded regions (cf. Fig. \ref{pic:stab:ana:FP}). The colored lines are derived
from the dispersion relations of the linearized hydrodynamic theory \eqref{eqn:hydro:theor:real} for different values of
the truncation parameter $N_c$, cf. Eq. \eqref{eqn:trunc:muD}. The critical line of the Turing instability (blue dashed
line
separating (A) and (B)) is not correctly predicted in the case $N_c=2$. Qualitative agreement is achieved for $N_c = 4$,
higher order approximations are almost exact. Parameters: $D_\varphi = 10$, others as in Fig. \ref{pic:stab:ana:FP}. }
   \label{pic:stab:ana:hydro:approxNC}
 \end{center}
\end{figure}
It is a priori not clear how many terms are necessary in order to reproduce quantitatively the instabilities found from
kinetic theory using a truncated operator
$\hat{\mu}_{\Delta,N_c}$. We close this paragraph by answering this question for our model. In the case of $N_c = 0$,
the results deduced from the spatially homogeneous equations are recovered (cf. Eq. \eqref{eqn:order:dis:trans:line}).
The results of the linear stability analysis for $N_c \in \left \{ 2,4,6,8 \right \}$
are shown in Fig. \ref{pic:stab:ana:hydro:approxNC}. It is evident from our analysis that it is insufficient to
approximate $\hat{\mu}_\Delta$ in the leading order $N_c = 2$. The disagreement is due to the fact that not all relevant
length scales are resolved properly in this case. However, the situation is essentially different if the approximation
$N_c = 4$ is used. This approximation enables the qualitative prediction of the structure of the phase space. Higher
order approximations $N_c \ge 6$ can be considered as almost exact on the linear level. We stress that the number of
terms which are necessary to reproduce the spectrum $\hat{\mu}_k$
sufficiently well, i.e. how many terms are necessary to resolve all relevant length scales, is not a universal number
but depends on model details. 

\subsection{Conclusions}

The analysis presented in this chapter reveals the existence of a Turing instability of the
coarse-grained velocity
field due to the anti-alignment interaction -- there exists a region in parameter space where the dynamics of the system
is linearly unstable at finite wavelengths. Furthermore, we showed that a system of hydrodynamic equations
\eqref{eqn:lin:hydrodyn:eqn:1545}, where the nonlocal operator $\hat{\mu}_{\Delta}$ is approximated by a truncated
series, provides an excellent approximation for the prediction of phase boundaries
given the following conditions are fulfilled:
\begin{itemize}
  \item The dynamics of the Fourier coefficients decouple, i.e. the truncation of the hierarchy of Fourier modes is
possible, if a considerable timescale separation among the modes exists. In
particular, high Fourier modes are strongly damped if angular fluctuations are large justifying the
negligence of these high order modes. This property allows to truncate the infinite hierarchy of Fourier modes and
enables us to replace it by a low dimensional system.
  \item The nonlocal interaction operator must be approximated sufficiently well, i.e. the series expansion
of $\hat{\mu}_{\Delta}$ must contain enough terms in order to resolve all relevant length scales. In our
particular model, this is the case if all terms including the term proportional to $\Delta^6 \, \vec{w}$ are taken
into account. 
 \end{itemize}
%
%
The linearized theory discussed here is not capable of predicting the macroscopic patterns since nonlinear terms were
not considered for simplicity. However, the theory confirms the existence of a Turing instability of the polar order
parameter in our model. Such an instability mechanism was proposed in \cite{dunkel_minimal_2013} on phenomenological grounds and 
shown to be sufficient to explain mesoscale turbulence in dense bacterial suspensions
\cite{wensink_meso-scale_2012,dunkel_fluid_2013}. In this work, we provide a microscopic explanation of this instability: the 
concurrence of short-range creation of local polar order with an opposing effect on larger scales.

\section{Summary}
\label{sec:sum}
 
We studied the pattern formation in an ensemble of self-propelled particles which move at constant speed and
interact via competing velocity-alignment interactions: at short length scales, particles tend to align their velocities
whereas they anti-align their direction of motion with particles which are further away \cite{grossmann_vortex_2014}.
The interplay of alignment and anti-alignment interaction has to be
understood as an effective description of two competing mechanisms: locally, polar order arises due to
anisotropic steric interactions or inelastic collisions for example, whereas the emergence of global polar order is
suppressed by an opposing effect, e.g. hydrodynamic interaction.  
 
Numerical simulations of the particle dynamics reveal the emergence of surprisingly complex
spatio-temporal pattern formation. On the one hand, our model reduces to a continuous time version of the Vicsek model
\cite{vicsek_novel_1995,peruani_mean_2008} for vanishing anti-alignment interaction. Therefore, we observe a
phenomenology similar to the Vicsek model \cite{chate_collective_2008} in this limiting case. Especially, large scale
traveling bands emerge in the low density regime close to the onset of polar order. 
Moreover, numerous additional phases were found due to the anti-alignment interaction. If this interaction is
strong and the particle density is high, we observe vortex arrays and synchronously rotating cluster. 
By gradually lowering the
anti-alignment interaction, the velocity-alignment acting on short length scales induces local polar order
and, thus, mesoscopic convective flows. In this intermediate regime, the anti-alignment
interaction inhibits the emergence of global polar order and, consequently, we observe a rotationally invariant state
possessing local polar order but global disorder (mesoscale turbulence). Another characteristic of this state is the emergence of transient 
vortices, which are separated by a characteristic
length. Moreover, we observe extended convective flows (\textit{jets}). At low densities, the actual patterns differ 
strongly: We observe small 
polar clusters moving
randomly for intermediate anti-alignment strength. At strong anti-alignment, the system
exhibits large scale nematic order as polar clusters start to form lanes moving in opposite directions.

Common to all the patterns is a characteristic length scale which manifests itself mainly in the momentum field.
Studying the linear stability of the disordered state within mean-field theory, we were able to relate
this pattern formation process to a Turing instability. Furthermore, we analyzed in
detail the possibility of describing this instability using a reduced set of hydrodynamic equations. We showed that a
system of hydrodynamic equations can quantitatively reproduce the relevant instabilities if two conditions are
fulfilled: (i) the reduction to a low dimensional system of order parameter equations is only possible if a considerable
timescale separation between fast and slow modes exists which is particularly fulfilled for high noise in our model;
(ii) the nonlocal interaction has to be described sufficiently well by taking into account a high number of derivatives
in the coarse-grained description. 


Based on the analysis of our self-propelled particle model, we suggest that the
interplay of competing alignment interactions may generally give rise to a Turing instability in the polar order
parameter field and that this mechanism in combination with convective particle transport can explain the emergence
of mesoscale turbulence and vortex arrays in active matter systems. 

\section*{Acknowledgments}

R. G. and M. B. acknowledge support by Deutsche Forschungsgemeinschaft (DFG)
through GRK 1558. 
P. R. acknowledges support by the German Academic Exchange Service (DAAD) through the P.R.I.M.E. Fellowship. 
L.S.G. thanks DFG for financial support through IRTG 1740. 

\appendix

\section{Mean-field approximation}
\label{sec:A:MFT}

The derivation of the nonlinear Fokker-Planck equation \eqref{eqn:nonl:FPE} relies on a mean-field
approximation (MFA)~--~the negligence of fluctuations~--~which we illustrate in detail in this section. The central
quantity is the microscopic particle density 
$\mathcal{P}(\vec{r},\varphi,t)$ defined by 
\begin{align}
  \mathcal{P}(\vec{r},\varphi,t) = \sum_{i=1}^N \delta^{(2)}(\vec{r} - \vec{r}_i(t)) \delta(\varphi - \varphi_i(t)) 
 \end{align}
which is a stochastic fluctuating field variable. From this field, we can compute the one-particle probability density
function to find a particle at position $\vec{r}$ with orientation $\varphi$ at time $t$ via
  \begin{align}
   P_1(\vec{r},\varphi,t) = \frac{\mean{\sum_{i=1}^N \delta^{(2)}(\vec{r} - \vec{r}_i(t)) \delta(\varphi -
\varphi_i(t))}}{N}   
     = \frac{\mean{\mathcal{P}(\vec{r},\varphi,t)}}{N}
  \end{align} 
as well as the two-particle probability density function to find one particle at position $\vec{r}$ with orientation
$\varphi$ and another particle at position $\vec{r}'$ with orientation $\varphi'$ at time $t$ via  \begin{subequations}
  \begin{align}
   P_2(\vec{r},\varphi;\vec{r}',\varphi';t)    
  &= \frac{\mean{ \sum_{i,j; i \neq j}
   \delta^{(2)}(\vec{r} - \vec{r}_i(t)) \delta^{(2)}(\vec{r}' - \vec{r}_j(t)) 
  \delta(\varphi - \varphi_i(t)) \delta(\varphi' - \varphi_j(t))}}{N(N-1)} \\
 &= \frac{\mean{\mathcal{P}(\vec{r},\varphi,t) \mathcal{P}(\vec{r}',\varphi',t)}}{N(N-1)} - \frac{P_1(\vec{r},\varphi,t)
\delta^{(2)}(\vec{r} - \vec{r}') \delta(\varphi - \varphi')}{N-1}.   \label{eqn:A:P2:exact}
  \end{align} 
 \end{subequations}
In mean-field approximation, fluctuating variables are approximated by their average values thus neglecting
fluctuations: 
 \begin{align}
  \mathcal{P}(\vec{r},\varphi,t) 
    &= \mean{\mathcal{P}(\vec{r},\varphi,t)} + \delta \mathcal{P}(\vec{r},\varphi,t) 
    \, {\underset{\text{(MFA)}}{\approx}} \, \mean{\mathcal{P}(\vec{r},\varphi,t)}.
 \end{align} 
Therefore, we deduce from \eqref{eqn:A:P2:exact} that we can approximate the two-particle probability density function
by the product of one-particle probability densities: 
\begin{subequations}
 \begin{align}
  P_2(\vec{r},\varphi;\vec{r}',\varphi';t) 
   & {\underset{\text{(MFA)}}{\approx}}  \frac{N P_1(\vec{r},\varphi,t) P_1(\vec{r}',\varphi',t)}{N-1} -
\frac{P_1(\vec{r},\varphi,t) \delta^{(2)}(\vec{r} - \vec{r}')\delta(\varphi - \varphi')}{N-1} \\
   & {\underset{(N \rightarrow \infty)}{\approx}} \, P_1(\vec{r},\varphi,t) P_1(\vec{r}',\varphi',t). 
 \end{align}
\end{subequations}
Thus, the mean-field approximation is equivalent to the negligence of inter-particle correlations. 


\bibliographystyle{MyEPJST}
\bibliography{SPAA.bib}

\end{document}